\documentclass[aps,prd,11pt,notitlepage]{revtex4}
    \usepackage{graphicx}
    \usepackage{amsmath,amssymb,mathrsfs}
\usepackage{epsf,bm}
\usepackage{slashed}
\usepackage{epsfig}
\usepackage[font=small,skip=0pt,justification=raggedright]{caption}

\newcounter{fig}   \newcommand{\lbfig}[1]{\refstepcounter{fig}
\label{#1} }

\newcommand{\bea}{\begin{eqnarray}}
\newcommand{\eea}{\end{eqnarray}}
\newcommand{\be}{\begin{equation}}
\newcommand{\ee}{\end{equation}}

\def\bfph{{\pmb{\phi}}}

\newcommand{\re}[1]{(\ref{#1})}

\begin{document}

\title{Soliton solutions of the fermion-Skyrmion system in (2+1) dimensions}

\author{I.~Perapechka}
\affiliation{Department of Theoretical Physics and Astrophysics,
Belarusian State University, Minsk 220004, Belarus}
\author{ Nobuyuki Sawado}
\affiliation{
Department of Physics, Tokyo University of Science, Noda, Chiba 278-8510, Japan
}
\author{Ya.~Shnir}
\affiliation{BLTP, JINR, Dubna 141980, Moscow Region, Russia
}
\begin{abstract}
We study effects of backreaction of the fermionic modes localized by the baby Skyrmion in the
(2+1)-dimensional Skyrme model.
It is shown that there is a tower of fermionic modes of two different types, localized by the soliton,
however there is only one fermionic level, which flows from positive to negative value as coupling
increases. Considering the strong coupling regime we observe that the coupling of the bosonic field to the
fermions may strongly deform the Skyrmion, in particular
the regions of negative topological charge density appear.
\end{abstract}
\maketitle

\section{Introduction}

The Skyrme model \cite{Skyrme:1961vq} is very well known as a prototype example of a field theory which supports
topological soliton solutions, the Skyrmions. The properties of Skyrmions were extensively studied over last decades,
for a review see, for example \cite{Manton:2004tk,Brown}. The standard (3+1)-dimensional Skyrme model
was can be considered as a non-linear low energy effective theory of Quantum Chromodynamics,
in the limit of a large number of quark colours \cite{Witten:1983tw}. In this framework the baryons
are considered as solitons of the effective  mean field theory, which can be constructed after integration over the
quarks, see e.g. \cite{Aitchison:1984ys,Dhar:1985gh,Ebert:1994mf}.
Then the pions correspond to the linearized fluctuations of the baryon field whereas the fermions, like quarks, do
not appear as fundamental physical fields.
Functional integration over the fermionic degrees of freedom yields so called chiral quark model
~\cite{Diakonov:1987ty,Wakamatsu:1990ud,Christov:1995vm}, an alternative description of baryons as
chiral solitons is given by the bosonized  Nambu-Jona-Lasinio model \cite{Alkofer:1994ph}.

A peculiar feature of the spectrum of the Dirac fermions
in the background of a Skyrmion is that, in agreement with the index theorem,
it shows a spectral flow of the eigenvalues with one normalizable bounded mode crosses zero
\cite{Hiller:1986ry,Kahana:1984dx,Kahana:1984be,Ripka:1985am}. Considering this problem, Kahana and Ripka
evaluated the contribution of the energy of quarks coupled to the Skyrmion
\cite{Kahana:1984dx,Kahana:1984be,Ripka:1985am}, it was shown that it does not destabilize the soliton.

The original Skyrme model shares many properties with its lower dimension planar analogue,
$\mbox{O}(3)$ non-linear sigma model, which is known as the baby Skyrme model \cite{BB,Leese:1989gj,Bsk1,Bsk}.
This simplified model finds various physical realizations, for example it
naturally arise in ferromagnetic structures with intrinsic and induced chirality \cite{Bogdanov}, or
in chiral nematic and anisotropic fluids \cite{Smalukh1,Smalukh2}. In such a context the baby Skyrme model is no
longer considered as an effective low energy theory, constructed via integration over the fermionic degrees of freedom.

Indeed, fermionic zero modes naturally appear in supersymmetric extensions of the $\mbox{O}(3)$ non-linear sigma model
\cite{Novikov:1984ac} and baby-Skyrme model \cite{Adam:2011hj,Queiruga:2016jqu}. Since there are four bosonic
collective coordinates of the charge one soliton solution of the $\mbox{O}(3)$ sigma model, there are 2 complex zero energy
fermionic states. In the baby Skyrme model the scale invariance is broken, so there are only 3 bosonic modes.
Correspondingly, the spinors possess three Grassmannian
degrees of freedom, the N=1 SUSY baby Skyrmion preserves only 1/4 supersymmetry \cite{Queiruga:2016jqu}.
We can expect that, as the supersymmetry becomes completely broken, the number of the fermionic zero
modes become further reduced.

Note that the fermionic fields coupled to the $\mbox{O}(3)$ non-linear sigma model in (2+1)-dimensions
were studied before in the context of the spin-statistic properties of the solitons. In this approach
the chiral field also appears as a result of an integration over fermions.  The leading term of the gradient expansion
yields the usual action of the $\mbox{O}(3)$ sigma model \cite{Jaroszewicz:1985ip, Abanov:2000ea,Abanov:2001iz}
supplemented by the topological $\theta$-term, which is proportional to the Hopf number $H_2$. This term is an analogue of the
Wess-Zumino-Witten term in the (3+1)-dimensional Skyrme model. As a result, the spin of the soliton becomes equal to a half of
the fermionic number. On the other hand, the Dirac fermions coupled to the baby Skyrmion  were considered an a
six-dimensional brane world model \cite{Kodama:2008xm, Delsate:2011aa}. However,
in these studies, no backreaction from the fermions was taken into account.

A main purpose of the present paper is to examine the effects of backreaction of the fermionic modes
coupled with baby-Skyrme model. Our numerical simulations confirm that, for a certain set of values
of other parameters of the model, there is one zero energy fermionic state.
In other words, there is just a single fermionic level, which flows from positive to negative value as coupling
decreases. This observation agrees with the statement of the index theorem, which sets a
correspondence between the number of zero modes and
the spectral flow of the fermionic Hamiltonian.

We found that similar to the case of the spherically symmetric (3+1)-dimensional fermions,
coupled  to the chiral field without backreaction, there are localized modes of different types.
Our numerical results reveal that, apart the nodeless quasizero mode, there are various shell-like modes,
which can be classified by the number of nodes $k$ of the fermionic field. We observe that
as a result of backreaction, the coupling to the fermionic modes may yield strong deformation of the Skyrmion, in
particular the regions of negative topological charge density appear. This effect can be interpreted as
production of a tightly bounded Skyrmion-anti-Skyrmion pair.

This paper is organised as follows. In Section II we present the planar Skyrme model,
coupled to the spin-isospin fermionic field. We restrict our consideration to simple rotationally invariant configuration of
topological degree one. Numerical results are presented in Section III, where we
describe the solutions of the model and discuss the energy spectrum of the localized fermionic states.
Conclusions and remarks are formulated in the last Section.

\section{The model}

The Lagrangian density of the coupled fermion-Skyrmion system in (2+1) dimensions can be written
as
\be
\label{lag}
\mathcal{L}=\mathcal{L}_{Sk}+\mathcal{L}_{f}.
\ee
where $\mathcal{L}_{Sk}$ is the Lagrangian of the planar Skyrme model
\cite{BB,Leese:1989gj,Bsk1,Bsk}
\be
\label{lagBS}
\mathcal{L}_{BS}=\frac{\kappa_2}{2}\left(\partial_\mu\bfph \right)^2-
\frac{\kappa_4}{4}\left(\partial_\mu\bfph \times \partial_\nu\bfph \right)^2-\kappa_0 V\, .
\ee
Here we are using the flat metric $g^{\mu\nu}=\mbox{diag}(1,-1,-1)$ and
the real triplet of scalar fields  $\bfph=(\phi_1,\phi_2,\phi_3)$ is constrained to
the surface of a sphere of unit radius:
$\bfph \cdot \bfph=1$.
The coupling constants $\kappa_0, \kappa_2$ and $\kappa_4$ are some real positive parameters. The
Lagrangian of the planar Skyrme model \re{lagBS} also includes a potential term $V$,
which, in the absence of other fields stabilizing the solitons. Here we consider the most common choice of the
$\mbox{O}(3)$ symmetry breaking potential
\be
\label{pot}
V=1-\phi_3 \, .
\ee

The topological restriction on the field $\bfph$ is that on the spatial boundary ${\bfph}_\infty = (0,0,1)$.
This allows a one-point compactification of
the domain space $\mathbb{R}^2$ to $S^2$ and
the field of the finite energy soliton solutions of the
model, so called baby Skyrmions, is a map $\bfph :S^2 \mapsto S^2$ which belongs
to an equivalence class characterized by the homotopy group $\pi_2(S^2)=\mathbb{Z}$.
The corresponding topological invariant is
\be \label{charge}
Q= - \frac{1}{4\pi}\int \bfph \cdot   (\partial_1 \bfph \times \partial_2 \bfph) ~d^2x
\ee

The Lagrangian for fermions coupled to the baby Skyrmions is given by
\be
\label{lagf}
\mathcal{L}_{f}=\bar{\Psi}\left(i \slashed{\partial}- g \pmb{\tau} \cdot \bfph - m \right)\Psi,
\ee
Note that the fermion field $\Psi$ is a spin and an isospin spinor, the isospin  matrices are defined as
$\pmb{\tau}=\mathbb{I}\otimes\pmb{\sigma}$, whereas the spin matrices are
$\hat{\gamma_\mu}=\gamma_\mu\otimes\mathbb{I}$. Here $\mathbb{I}$ is two-dimensional
identity matrix, $\slashed{\partial}=\hat{\gamma_\mu}\partial^\mu$ and $\pmb{\sigma}$ are
the usual Pauli matrices. The coupling constant $g$ and the fermions mass  $m$
are parameters of the model.
In order to satisfy the usual anticommutation relations of the Clifford algebra, the two-dimensional
gamma-matrices are defined as follows:
$\gamma_1=-i\sigma_1, \gamma_2=-i\sigma_2$ and $\gamma_3=\sigma_3$.

There are five parameters of the coupled fermion-Skyrmion system \re{lag},
$\kappa_0, \kappa_2, \kappa_4, g$ and $m$. The dimensions of the parameters and the fields are
\be
\begin{split}
\kappa_0&: ~\left[L^{-3}\right]\, , \quad
\kappa_2: ~\left[L^{-1}\right]\, , \quad \kappa_4: ~\left[L^1\right]\, , \quad
g: ~\left[L^{-1}\right]\\
m&:~\left[L^{-1}\right]\, , \quad \bfph:~\left[L^{0}\right]\, , \quad
\Psi: ~\left[L^{-1}\right] \, . \nonumber
\end{split}
\ee
Thus, an appropriate rescaling of the action by an overall
constant, and rescaling of the length scale and the fermion field
\be
\label{rscale}
r\to\sqrt\frac{\kappa_4}{\kappa_2}r,\quad
\Psi\to\sqrt{\frac{\kappa_2}{\kappa_4}} \Psi, \quad
\kappa_0\to\frac{\kappa_2^2}{\kappa_4}\kappa_0, \quad
g\to\sqrt\frac{\kappa_2}{\kappa_4}g, \quad m\to\sqrt\frac{\kappa_2}{\kappa_4}m \, ,
\ee
effectively reduces the number of independent parameters to four,
$\sqrt{\kappa_2\kappa_4}, \kappa_0, m$ and $g$. Hereafter we fix $\sqrt{\kappa_2\kappa_4}=1$ without loss of generality.

Considering the model \re{lag} we do not impose the usual assumption that
the Skyrmion is a fixed static background field \cite{Hiller:1986ry,Zhao:1988in,Balachandran:1998zq,Krusch:2003xh}.
We restrict our consideration to the rotationally invariant stationary configuration of topological degree $Q=1$, thus
$\Psi=\psi e^{-i\varepsilon t}$ and we suppose that the Skyrmion field $\bfph$ is static.
Then the rescaled Hamiltonian of the coupled system can be written as:

\be
\label{energy}
H=\int{d^2x\;\psi^{\dagger}\mathcal{H}\psi} + \sqrt{\kappa_2\kappa_4} \int{d^2x\left(\frac{1}{2}\left(\partial_i\bfph
\right)^2+\frac{1}{4}\left(\partial_i\bfph  \times \partial_j \bfph  \right)^2+\kappa_0 V\right)} \, ,
\ee
where
\be
\label{ham}
\mathcal{H}=\hat{\gamma_3}\left(-i\hat{\gamma_k}\partial_k+ g \pmb{\tau}\cdot\bfph  +m\right)
\ee
is the fermionic Hamiltonian.

Variation of the action of the model \re{lag} with respect to the
fermion field $\bar \psi$ yields the Dirac equation
\be
\label{Dirac}
\mathcal{H}\psi^{(i)}=\varepsilon^{(i)} \psi^{(i)} \, ,
\ee
with eigenvalues $\varepsilon^{(i)}$. Here the superscript $i$ corresponds to a particular fermionic level
(with no sum on repeated indices in \re{Dirac}).

Apart the topological density of the Skyrme field \re{charge} we also consider
the fermionic density
\be
\label{probch}
\rho=\bar{\psi}\hat{\gamma_3}\psi=\psi^{\dagger}\psi \, .
\ee
As we will see there are fermionic configurations localized by the Skyrmion.

The field equation for the Skyrme field can be conveniently written
in the form
\be
\partial_\mu j^\mu = \kappa_0 {\bfph}_\infty \times \bfph + \frac{g}{\sqrt{\kappa_2\kappa_4}} \bfph \times \left(\psi^\dagger
\pmb{\tau}  \psi \right) \, ,
\label{eqsSk}
\ee
where the scalar current is defined as \cite{BB,Leese:1989gj,Bsk}
\be
j_\mu = \bfph \times \partial_\mu \bfph + \partial_\nu \bfph \left(
\partial^\nu \bfph \cdot (\bfph \times \partial_\mu \bfph)
\right)
\label{current}
\ee

\subsection{Rotationally invariant configurations}

Thereafter we consider simple rotationally invariant baby Skyrmion, which is parametrized by the
ansatz:
\be
\label{ansSk}
\bfph=\left(\sin f(r) \cos n\varphi, \sin f(r) \sin n\varphi, \cos f(r)\right) \, .
\ee
Here $f(r)$ is some monotonically decreasing
radial function, $n\in \mathbb{Z}$  and $\varphi$ is the usual azimuthal angle. Since the field must approach the vacuum on the spacial
asymptotic, it satisfies the boundary condition $\cos f(r) \to 1$ as $r \to \infty$, i.e., $f(\infty) \to 0$.

Further, the fermionic Hamiltonian can be written explicitly in the matrix form as
\be
\label{hamdens}
\mathcal{H}=\left(
\begin{array}{cccc}
g \cos f +m & g e^{-i n\varphi}\sin f & -e^{-i\varphi}\left(\partial_r-\frac{i\partial_\varphi}{r}\right) & 0\\
g e^{i n\varphi}\sin f & -g \cos f +m & 0 & -e^{-i\varphi}\left(\partial_r-\frac{i\partial_\varphi}{r}\right) \\
e^{i\varphi}\left(\partial_r+\frac{i\partial_\varphi}{r}\right) & 0 & -g \cos f -m & -g e^{-i n\varphi}\sin f \\
0 & e^{i\varphi}\left(\partial_r+\frac{i\partial_\varphi}{r}\right) & -g e^{i n\varphi}\sin f & g \cos f -m
\end{array}
\right)\, .
\ee
The corresponding rotationally invariant spin-isospin eigenfunctions with the eigenvalues $\epsilon^{(i)}$
can be written as
\be
\label{ansFer}
\psi^{(i)}= \mathcal{N}^{(i)} \left(
\begin{array}{c}
v_1 e^{i l \varphi}\\
v_2 e^{i (l+n) \varphi}\\
u_1 e^{i (l+1) \varphi}\\
u_2 e^{i (l+n+1) \varphi}
\end{array}
\right)\, ,
\ee
where spinor components $u_i$ and $v_i$ are
functions of radial coordinate only, $l\in\mathbb{Z}$ and $\mathcal{N}^{(i)}$ is a normalization factor, which is defined
from the usual condition
\be
\int d^2x ~{\psi^{(i)}}^\dagger \psi^{(i)}
=2\pi {\mathcal{N}^{(i)}}^2 \int_0^\infty rdr (v_1^2+v_2^2+u_1^2+u_2^2)=1\,.
\label{norm}
\ee
We restrict our consideration below to the fermionic states with filling factor one.

The rotationally invariant fermionic Hamiltonian \re{hamdens} commutes
with the total angular momentum operator
\be
\label{grandspin1}
K_3=-i\frac{\partial}{\partial\varphi}+\frac{\hat{\gamma_3}}{2}+n\frac{\tau_3}{2}\, .
\ee
The corresponding half-integer eigenvalues $\kappa=\frac{1}{2}\left(1+n+2l\right)$ can be used together with the
topological charge of the soliton to
classify different field configurations. The ground state corresponds to $\kappa=0$ and thus $l=-\frac{1}{2}(1+n)$ and
in the ground state $l=0$ as $n=-1$.

It is instructive to investigate the asymptotic behavior of the fields at spatial infinity.
Then the Skyrme field is approaching the vacuum, $\bfph \approx {\bfph}_\infty + \delta \bfph$,
where $\delta \bfph \cdot {\bfph}_\infty =0$.
Thus, the asymptotic expansion of the equations \re{Dirac},\re{eqsSk} at $r\to \infty$ yields two decoupled
linearized equations
\be
\begin{split}
(-i \hat{\gamma_k}\partial_k + g\tau_3 +m  )\psi &=0\, , \\
(\Delta - \kappa_0  )\delta \bfph &= 0 \, .
\label{eqasymptot}
\end{split}
\ee
As is well known \cite{Bsk,PZS}, the asymptotic solution for the rotationally symmetric
field of the Skyrmion of
topological degree $Q=n$ is given by the modified Bessel function
\be
\delta \bfph \sim  K_n\left(\sqrt{\kappa_0} r) (\cos\left(n\varphi - \chi\right) , \cos\left(n\varphi - \chi\right) , 0\right)\, ,
\label{bessel}
\ee
where the angle $\chi$ corresponds to the orientation of the Skyrmion. Thus, the
soliton is exponentially localized
and the asymptotic field $\delta \bfph $ may be thought of as generated by a pair of orthogonal $2^n$-poles.

Consequently, the first asymptotic equation \re{eqasymptot}
on the  fermionic spin isospin spinor $\psi=\left(v_1,v_2,u_1,u_2\right)$
can be expressed in components in the form of two pairs of identical second order equations:
\be
\begin{split}
(\Delta - 4(g \pm m )^2  )u_{1,2} &= \varepsilon^2u_{1,2} \, ;\\
(\Delta - 4(g \pm m )^2  )v_{1,2} &= \varepsilon^2v_{1,2} \, .
\end{split}
\label{eqsspinorcomp}
\ee
Thus, the components of the rotationally symmetric fermionic field decay as
\be
\begin{split}
v_1 &\sim e^{il(\varphi - \chi)} K_l(\sqrt{4(g+m)^2- \varepsilon^2}\, r)\, ,\\
v_2 &\sim e^{i(l+n)(\varphi - \chi)} K_{l+n}(\sqrt{4(g-m)^2- \varepsilon^2}\, r) \, ,
\end{split}
\ee
and similar for the components $u_{1,2}$. Evidently, the real and imaginary parts of these
components are of the form \re{bessel}. In other words, for continuous band of eigenvalues
$|\varepsilon|<g-m$ there are fermionic fields
exponentially localized on the Skyrmion. Further, the fermion field
asymptotically represents a pair of orthogonal
$2^l$-poles, together with a pair of collinear $2^{l+n}$-poles.

Substitution of the ansatz \re{ansSk} and \re{ansFer} into the action of the coupled model \re{lag}
after some algebra yields the system of variational equations
\be
\label{eqs}
\begin{split}
\left(r +\frac{n^2 \sin^2 f}{r}\right)f''&+\frac{n^2\sin 2f}{2r}f'^2+\left(1-\frac{n^2 \sin^2 f}{r^2}\right)f'
-\frac{n^2\sin 2f}{2r} - \kappa_0 r \sin f \\&
+ \frac{g r \mathcal{ N}^2}{\sqrt{\kappa_2\kappa_4}}\left( \sin f (v_1^2+u_2^2-u_1^2-v_2^2)
+2 \cos f (u_1 u_2-v_1 v_2) \right)=0,\\
u_1'&+\frac{l+1}{r}u_1-g\sin f v_2+\left(\varepsilon-g\cos f-m\right)v_1=0,\\
u_2'&+\frac{l+n+1}{r}u_2-g\sin f v_1+\left(\varepsilon+g\cos f-m\right)v_2=0,\\
v_1'&-\frac{l}{r}v_1-g\sin f u_2-\left(\varepsilon+g\cos f+m\right)u_1=0,\\
v_2'&-\frac{l+n}{r}v_2-g\sin f u_1-\left(\varepsilon-g\cos f+m\right)u_2=0.
\end{split}
\ee
Solutions of these equations give the symmetric stationary point of the total energy functional.

The equations \re{eqs}, together with constraint imposed by the normalization condition \re{norm},
yields a system of integro-differential equations, which
can be solved numerically as we impose appropriate boundary conditions.
As usual, they follow from the conditions of regularity of the fields both
at the origin and at the spatial boundary,
and condition of finiteness of the energy of the system. In particular we have to take into account that
the asymptotic value of the scalar field is restricted to the vacuum and we are looking for localized solutions.

Regularity at origin leads to the following restrictions on the fields:
\be
\label{regor}
f\bigl.\bigr|_{r=0}=q\pi,\quad l v_1\bigl.\bigr|_{r=0}=\left(l+n\right)
v_2\bigl.\bigr|_{r=0}=\left(l+1\right) u_1\bigl.\bigr|_{r=0}=\left(l+n+1\right) u_2\bigl.\bigr|_{r=0}=0\, ,
\ee
where $q\in \mathbb{Z}$. Regularity at the spacial boundary and the condition of localization
of the fermionic field yields
\be
\label{reginf}
f\bigl.\bigr|_{r\rightarrow\infty}=v_1\bigl.\bigr|_{r\rightarrow\infty}=v_2\bigl.\bigr|_{r\rightarrow\infty}=
u_1\bigl.\bigr|_{r\rightarrow\infty}=u_2\bigl.\bigr|_{r\rightarrow\infty}=0\, .
\ee

Evidently, with these conditions the topological charge \re{charge} of the baby Skyrmion becomes
\be
\label{topbc}
Q=\frac{n}{2}\left((-1)^q -1\right) \, .
\ee
However, it is well known that higher charge planar Skyrmions generally
do not possess rotational invariance \cite{PZS,Bsk,Weidig:1998ii}.
Therefore we will restrict our consideration below
to the case $n=-1$, $q=1$, i.e. $Q=1$. Note that for the anti-Skyrmion with topological charge $Q=-1$ we have
$n=q=1$ and, in the ground state $l=-1$. Thus, the corresponding system of the field equations for the fermions, localized
on anti-Skyrmion, is identical with
\re{eqs} up to replacements $u_i \to v_i$,  $n \to -n$ and $\epsilon \to -\epsilon$,
i.e. the isospin reflection swaps the components of the fermionic field and inverts the sign of the fermion
energy.

\section{Numerical results}
To solve the system of of integro-differential equations \re{eqs} with constraint \re{norm}
numerically, we restrict
the radial variable $r$ to a compact interval $x=\frac{r}{1+r}$, so $x\in [0,1]$.
The  system is solved iteratively using Newton-Raphson method,
based on 6\textsuperscript{th} order central finite difference scheme.
The resulting system of linear algebraic equations is solved with direct PARDISO solver \cite{pardiso}.
Method is implemented in Wolfram Language.
The simulations use a grid size of  $N=1000$ nodes,
selected runs were repeated with other values of  $N$ to check the stability of our results.
The relative errors of calculations are lower than $10^{-8}$.

\begin{figure}[t]
\begin{center}
\includegraphics[width=.7\textwidth, trim = 60 20 80 50, clip = true]{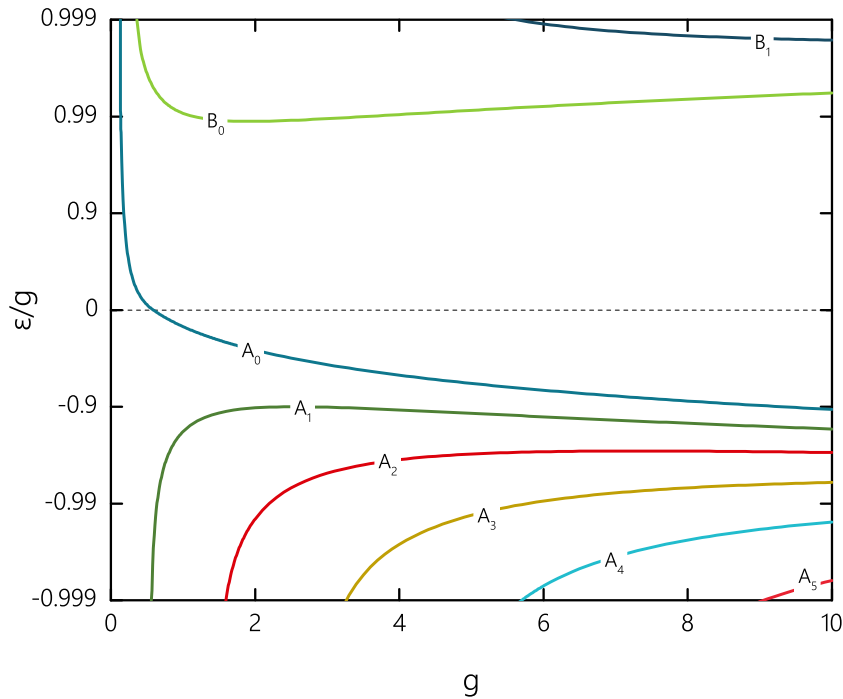}
\end{center}
\caption{\small
Normalized energy  $\frac{\varepsilon}{g}$ of the localized fermionic states
as a function of the fermion-Skyrmion coupling $g$
for several fermion modes at $m=0$ and $\kappa_0=0.1$.}
\lbfig{modeslim}
\end{figure}

\begin{figure}[t]
\begin{center}
\includegraphics[width=.45\textwidth, trim = 40 20 90 20, clip = true]{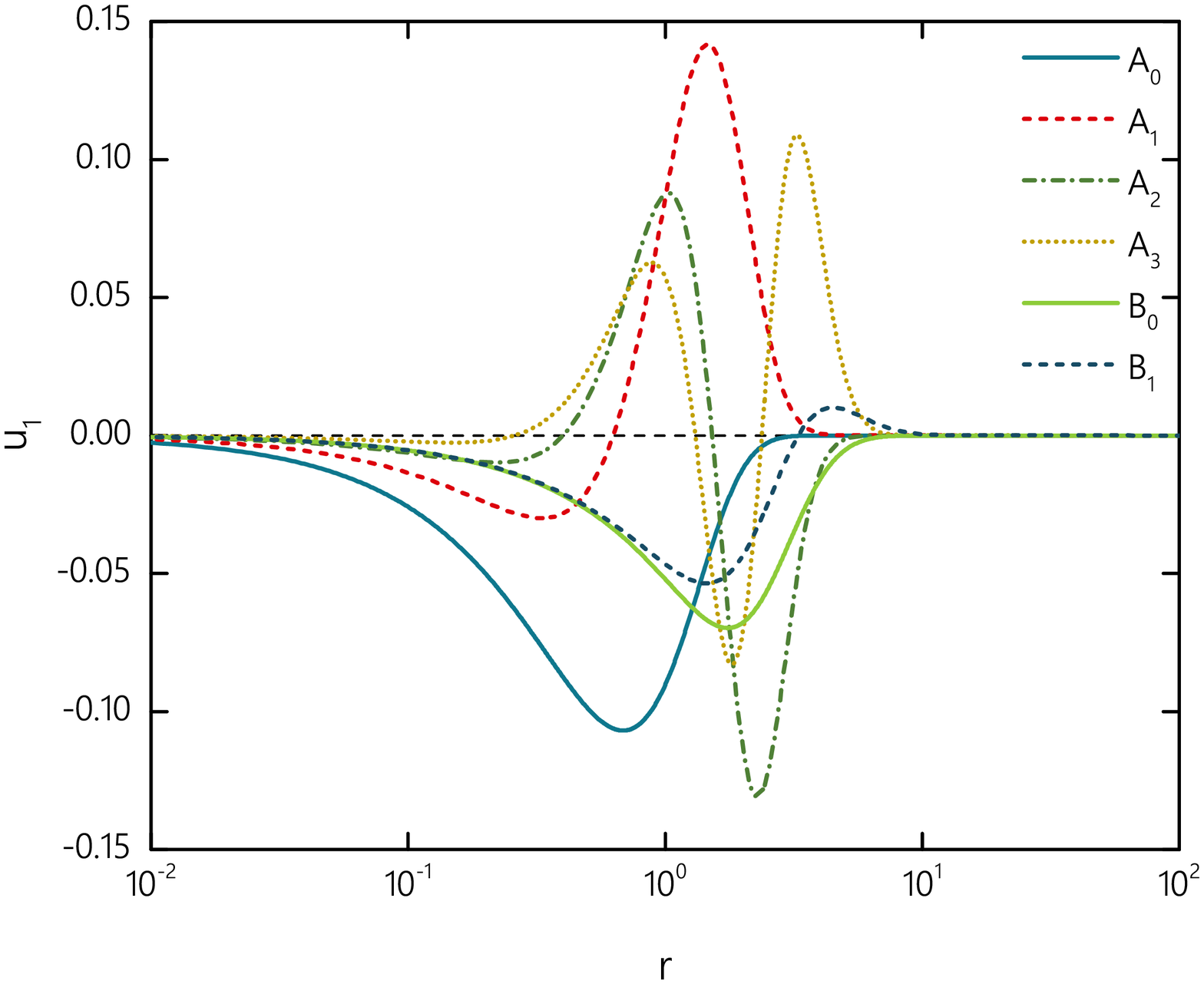}
\includegraphics[width=.45\textwidth, trim = 40 20 90 20, clip = true]{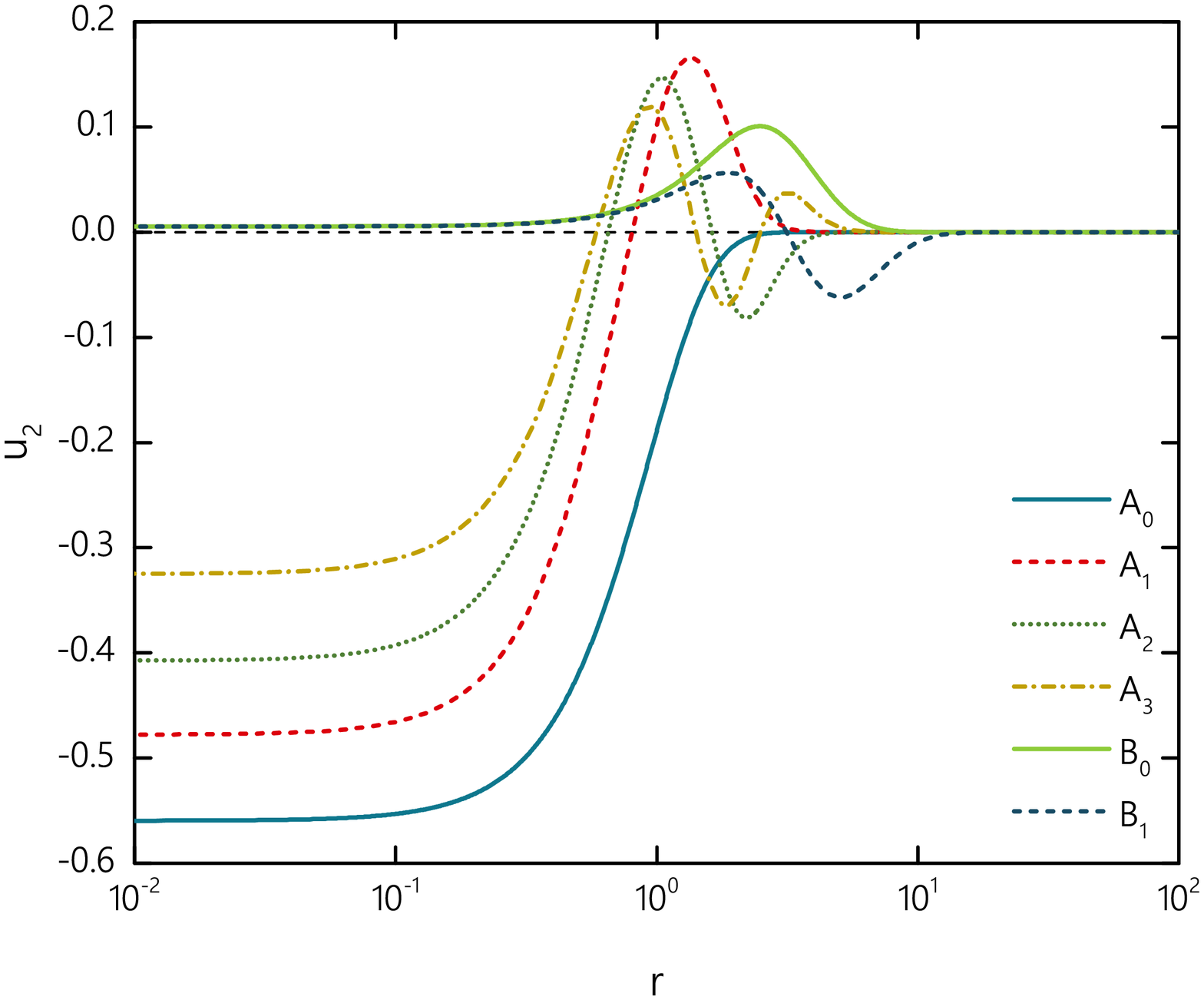}
\includegraphics[width=.45\textwidth, trim = 40 20 90 20, clip = true]{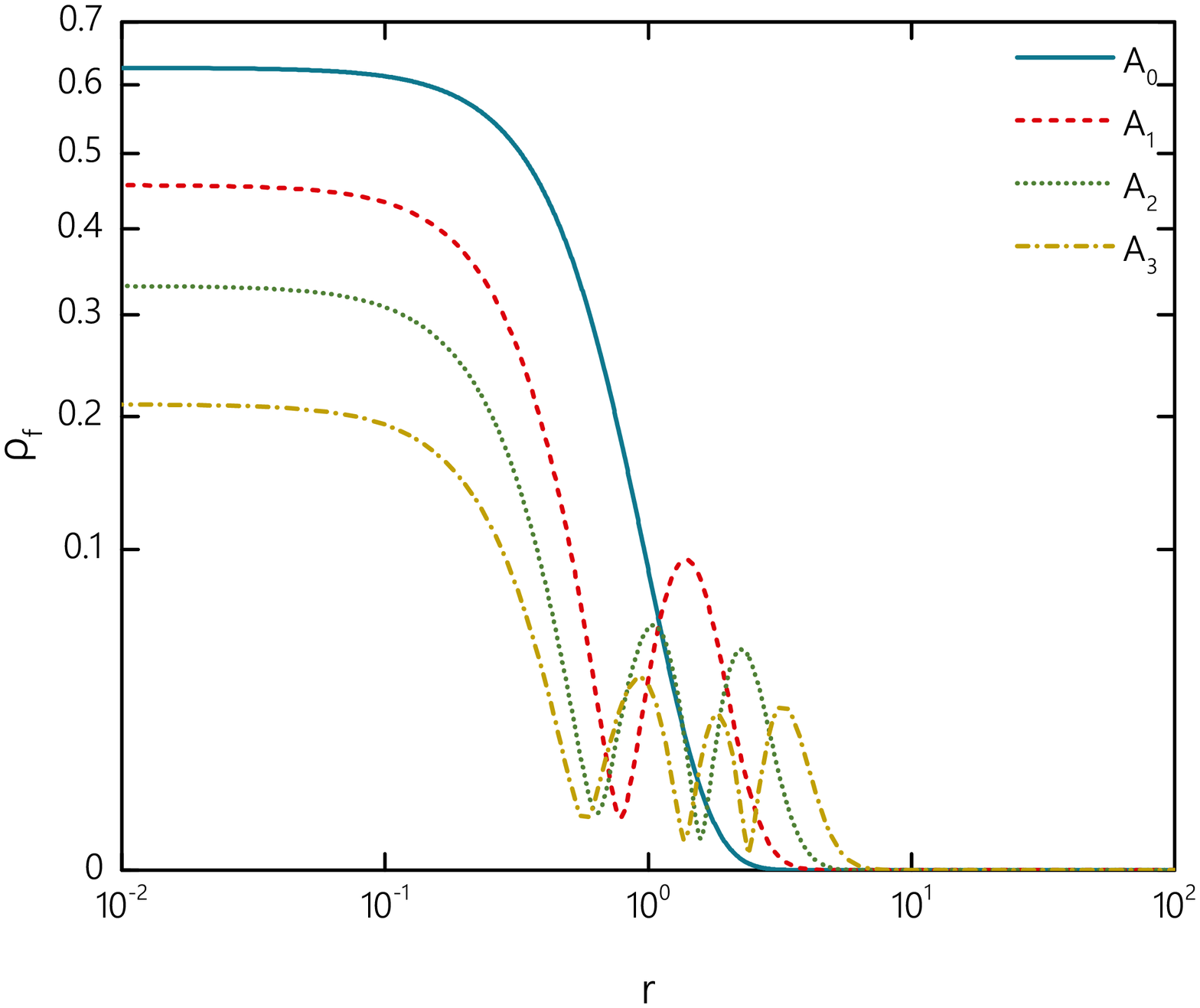}
\includegraphics[width=.45\textwidth, trim = 40 20 90 20, clip = true]{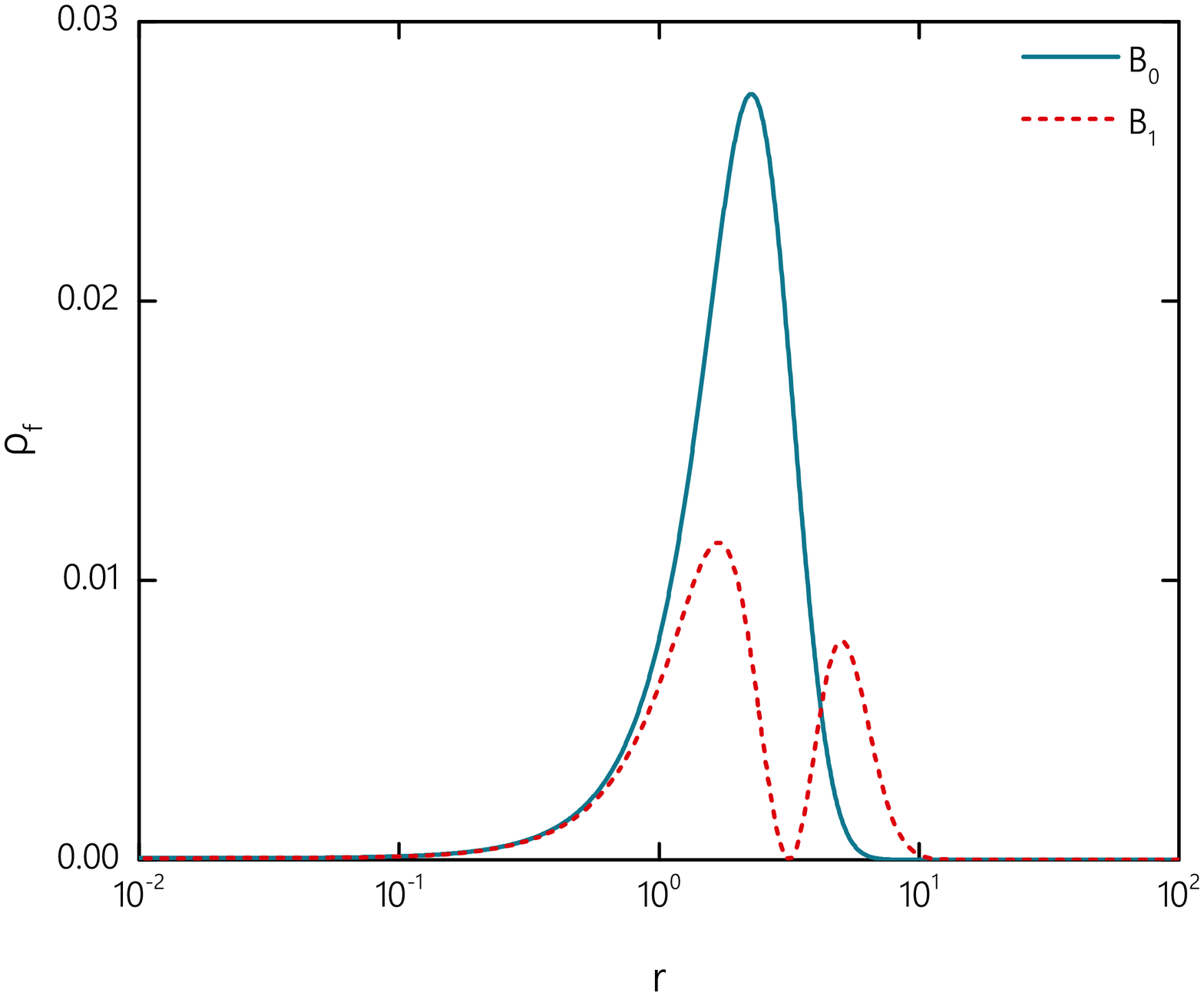}
\end{center}
\caption{\small
Fermionic field profile functions $v_1=u_2$ (upper left) and
$v_2=-u_1$ (upper right) and the fermionic densities for $A_k$ modes $\rho_A$ (lower left)
and for $B_k$ modes $\rho_B$ (lower right) are plotted as functions of the radial coordinate
$r$ at $g=1$, $m=0$ and $\kappa_0=0.1$.}
\lbfig{modes}
\end{figure}

Let us now restrict our consideration to the rotationally invariant
fermionic ground state, thus we fix $l=0$. Then the boundary conditions at the origin \re{regor} become simple
\be
\label{bcor}
f\bigl.\bigr|_{r=0}=\pi,\quad v_1\bigl.\bigr|_{r=0}=v_2'\bigl.\bigr|_{r=0}=u_1'\bigl.\bigr|_{r=0}
=u_2\bigl.\bigr|_{r=0}=0 \, ,
\ee
while the asymptotic boundary conditions \re{reginf} remain the same. Note that for the massless fermions
the ground state becomes accidentally degenerated, since in that case $u_2=v_1$ and $u_1=-v_2$.



The energy spectrum $\varepsilon$ of the spin isospin fermions coupled to
the background field of the Skyrmion was considered in several papers
\cite{Hiller:1986ry,Zhao:1988in,Balachandran:1998zq,Krusch:2003xh}. In this approximation
the profile function $f(r)$ of the Skyrmion field is not affected by the coupling strength $g$.
It corresponds to the case of weak coupling limit of our numerical simulations.
The pattern becomes different as
we take into consideration the backreaction of the strongly coupled localized fermions.
Indeed, for a finite value of the coupling strength $g$, the profile of the Skyrme field deforms as
the fermion occupied an energy level, further, the
energy levels move accordingly. Our numerical calculations continue until the self-consistency is attained.
In this sense, as the coupling $g$ grows,
we obtain an infinite tower of new Skyrmion solutions corresponding to the different filling factors and types
of the fermions occupying the energy levels.

\begin{figure}[t]
\begin{center}
\includegraphics[width=.7\textwidth, trim = 40 20 90 20, clip = true]{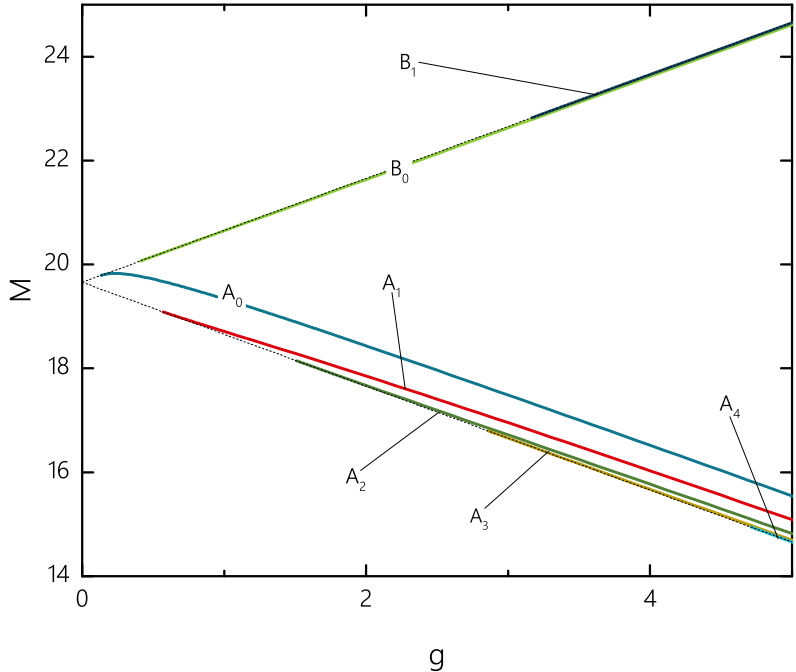}
\end{center}
\caption{\small
The total energy (\ref{energy}) for changing the coupling constant $g$.
}
\lbfig{Mtotalg}
\end{figure}

In Fig.~\ref{modeslim} we display the fermion energy (in unit of $g$) as a function of the coupling strength $g$.
We observe one zero-crossing mode from positive to negative continuum, which corresponds to zero mode supported by the index theorem.
The mode is found in previous studies of without backreaction: \cite{Kahana:1984dx,Kahana:1984be,Ripka:1985am}
shows the discrete energy spectrum of the
spherically symmetric (3+1)-dimensional fermions, coupled to the chiral field without backreaction. Further,
in the latter case there also are localized modes of different types,
which are counterparts of the modes of the types $A$ and $B$, respectively.

Considering the fermion modes localized by the Skyrmion with backreaction, we find that
there are two groups of the modes showing different behaviors,
which we shall refer to as $A_k$-modes and $B_k$-modes ($k=1,2,3,\ldots$), respectively.
Our numerical results reveal that for growing the coupling constant $g$, there appears an
infinite tower of the fermionic excitations localizing on the Skyrmion.
The eigenvalues of the $B_k$-modes are always positive, and are always close to the positive continuum threshold (see
Fig.~\ref{modeslim}). Similarly, the eigenvalues of the excited modes $A_k$, $k\ge 1$ are negative,
these modes are linked to the negative energy continuum (Dirac sea) approaching
it at some set of critical values of the fermion-Skyrmion coupling $g$.

In order to clearly see the difference, in Fig.~\ref{modes}, we plot the components of the wave function $v_1,v_2$
of the $A_k, B_k$ and also the fermionic densities.
As can be seen, the fermionic density distribution of the $A_k$ possesses a maximum at the center of the Skyrmion core, it
monotonically reduces to zero at infinity. The fermionic density distributions of the $B_k$
are vanishing both at the center of the Skyrmion and at infinity, is featuring an annular shape.
Apart the nodeless mode $A_0$, the shell-like modes of both types somewhat resemble the Bartnik-McKinnon solutions
in the Einstein-Yang-Mills theory \cite{Bartnik:1988am}, in which the solutions are
classified by number of nodes $k$ of the fields.
The difference between the modes of the type $A$ and $B$ is related with their decoupling limit;
the $A_k$-modes are linked to the Dirac sea while the $B_k$ modes emerge from the positive continuum.
Later we shall discuss this pattern in more detail.

We observe that for a certain set of values of the parameters of the model,
there is only one zero energy fermionic state.
It corresponds to the mode of type $A_0$ with eigenvalue $\varepsilon=0$.
In other words, there is just a single fermionic level,
which flows from positive to negative value as coupling $g$ decreases, see
In Fig.~\ref{modeslim}. This agrees with the statement of the index theorem, which sets a correspondence between the number of zero modes and
the spectral flow of the fermionic Hamiltonian \re{hamdens}.

We evaluate the total energy  (\ref{energy}) with the ansatz (\ref{ansSk}),(\ref{ansFer})
for the $A_k$-modes and $B_k$-modes.
Fig.\ref{Mtotalg} plots for the $A_k,k=0,\ldots,4$ and $B_k,k=0,1$.
Again we confirm that the $A_k$ and $B_k$ modes  behave quite differently, as
the coupling constant grows, all modes of type $A_k$ become strongly bounded to the Skyrmion decreasing the total energy of
the system. Oppositely, coupling to the modes of type $B_k$ increase the total energy of the bounded system.

In Fig.~\ref{fermodes}, we plot the fermionic density distributions \re{probch}
of the first three localizing modes $A_0,A_1,A_2$ for several values of coupling constant $g:10\leq g\leq 1000$.
At small values of the coupling constant $g$ there is only one localizing mode $A_0$ which should exists according to
the index theorem. As the coupling  $g$ increases, the effects of the backreaction becomes more visible, also the
higher $A_k$ modes become localized by the Skyrmion.
Further increase of the coupling $g$ yields stronger bounding of the modes, also larger number of localized
modes are extracted from the positive and negative continuum.

\begin{figure}[t]
\begin{center}
\includegraphics[width=.48\textwidth, trim = 40 20 90 20, clip = true]{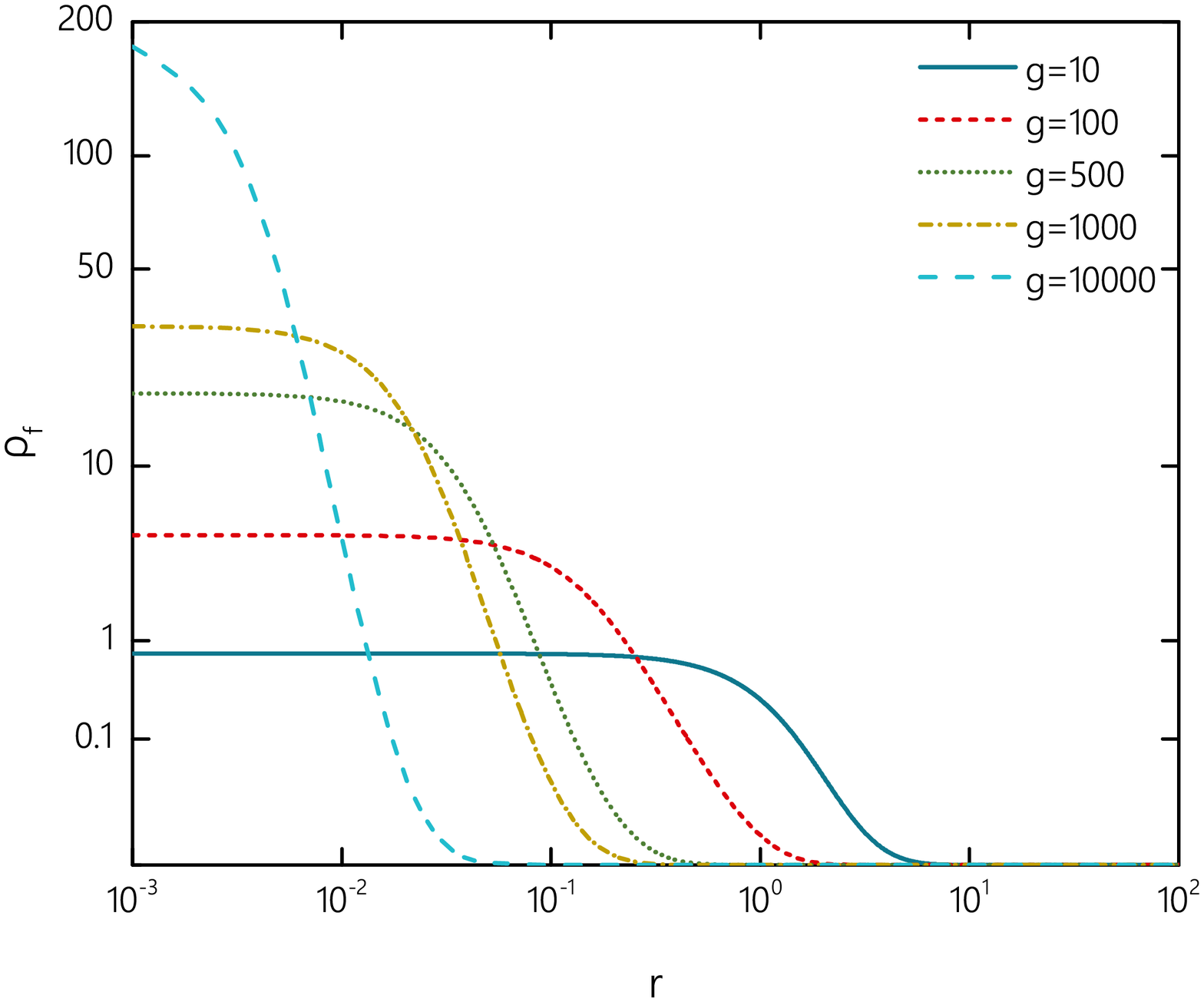}
\includegraphics[width=.48\textwidth, trim = 40 20 90 20, clip = true]{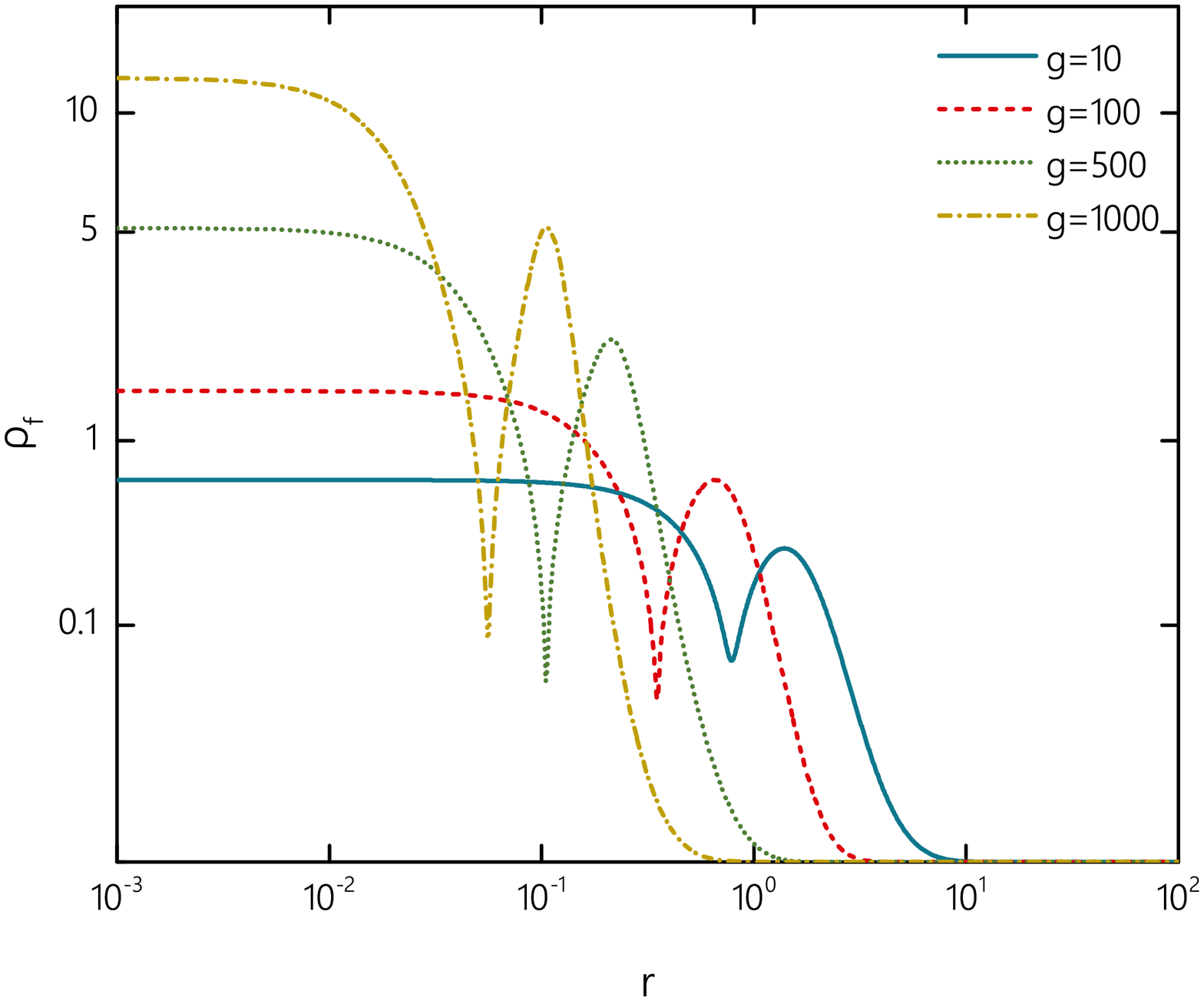}
\includegraphics[width=.48\textwidth, trim = 40 20 90 20, clip = true]{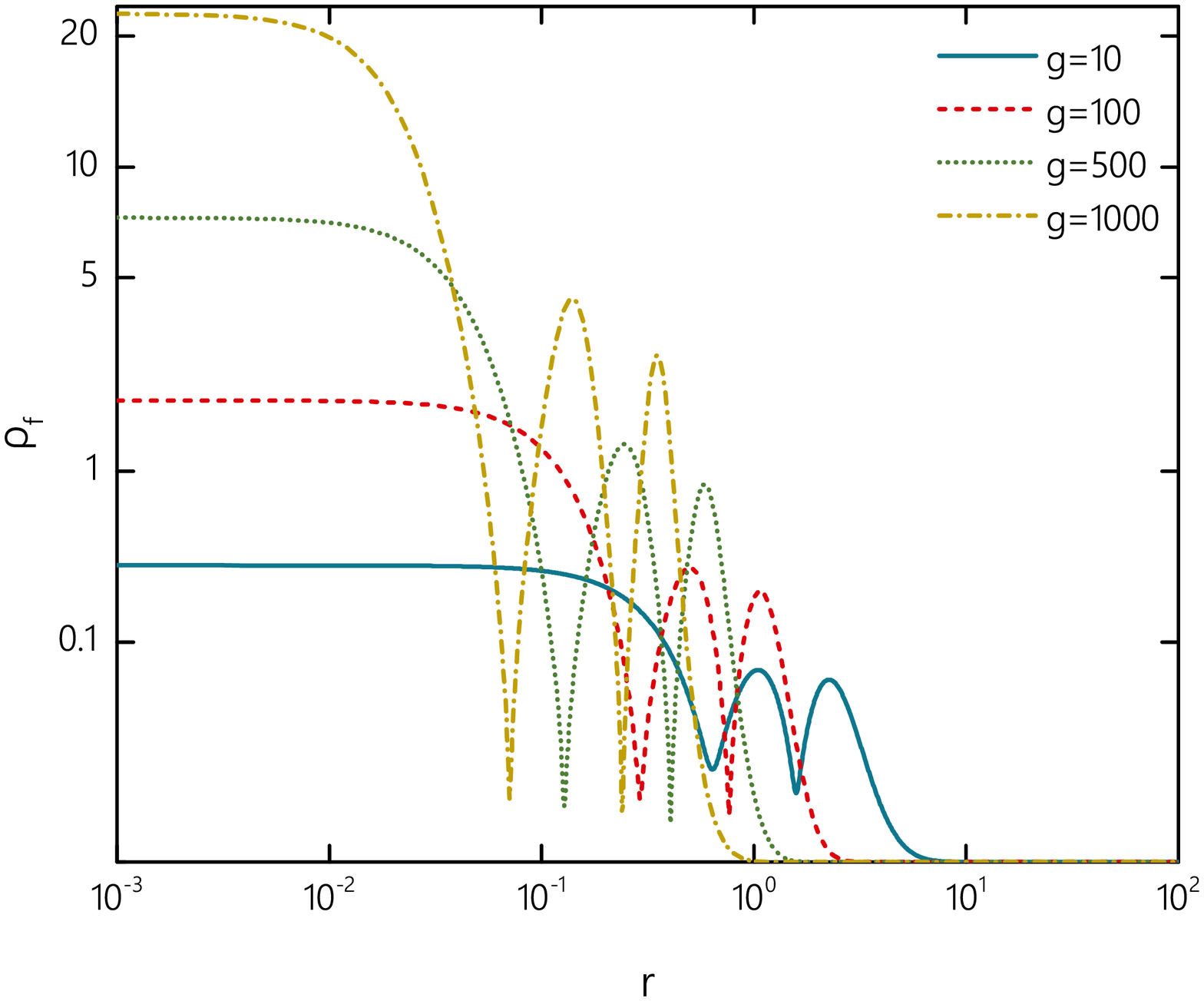}
\end{center}
\caption{\small
Fermionic density distributions of the localizing modes $A_0$ (upper left plot), $A_1$ (upper right plot) and $A_2$ (bottom plot)
as functions of the radial coordinate $r$ for $m=0$, $\kappa_0=0.1$ and several values of coupling constant $g:10\leq g \leq 100$.
}
\lbfig{fermodes}
\end{figure}

\begin{figure}[hbt]
\begin{center}
\includegraphics[width=.7\textwidth, trim = 60 20 80 50, clip = true]{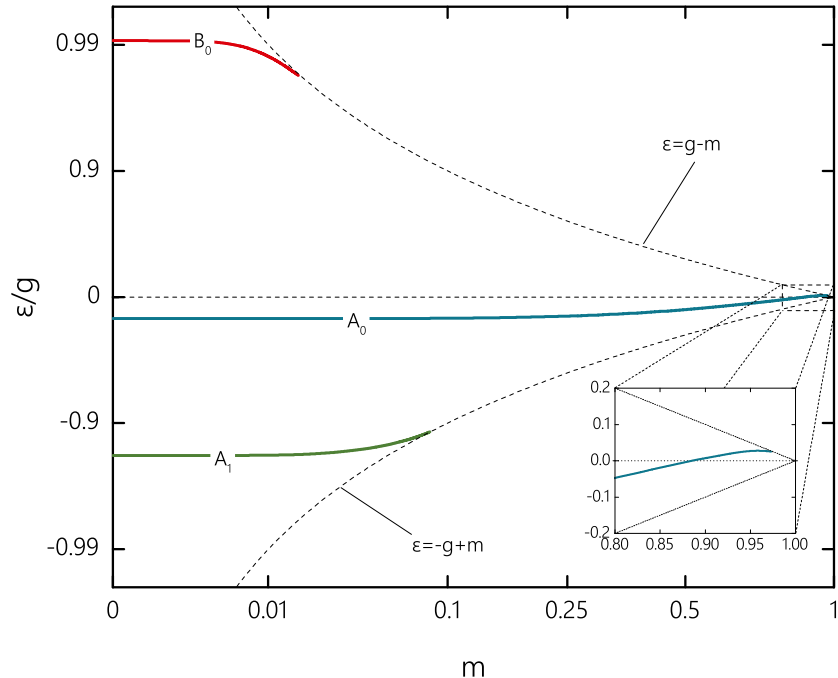}
\end{center}
\caption{\small
Normalized energy  $\frac{\varepsilon}{g}$ of the localized fermionic states
as a function of the fermion mass $m$
for several fermion modes at $g=1$ and $\kappa_0=0.1$. The dashed lines indicate positive and negative continuum thresholds
$\varepsilon =\pm(g-m)$.}
\lbfig{modesm}
\end{figure}

\begin{figure}[t]
\begin{center}
\includegraphics[width=.45\textwidth, trim = 40 20 90 20, clip = true]{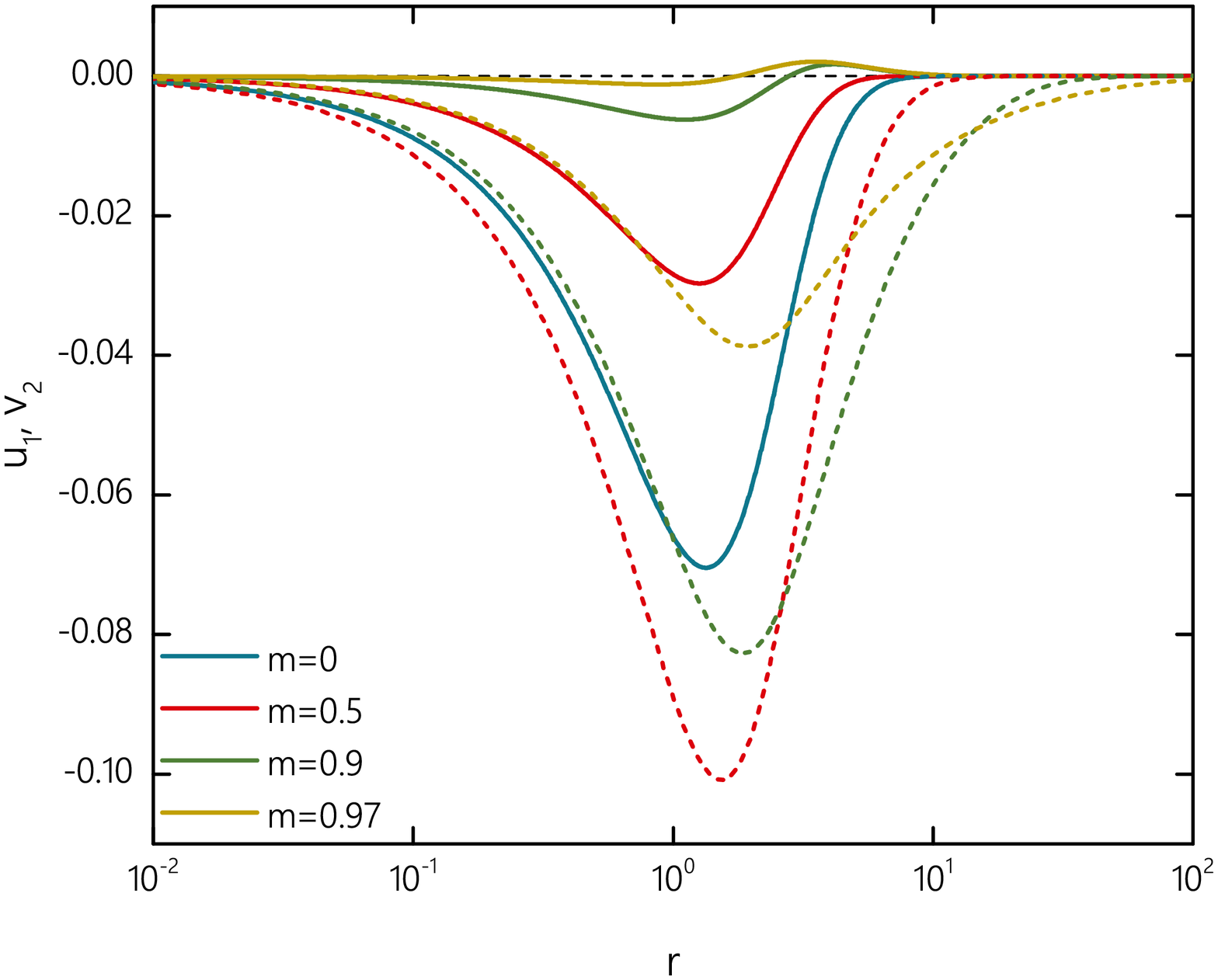}
\includegraphics[width=.45\textwidth, trim = 40 20 90 20, clip = true]{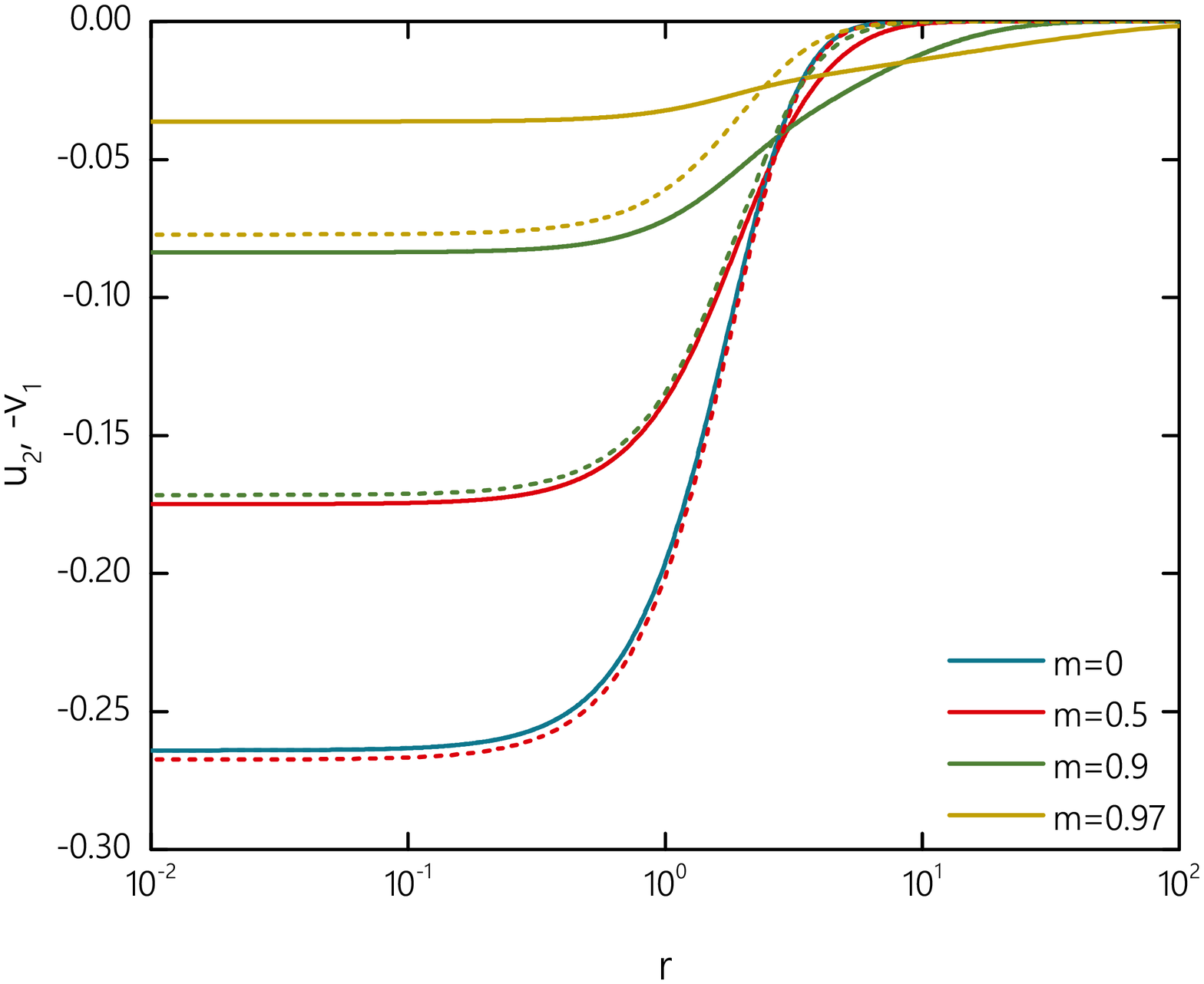}
\includegraphics[width=.45\textwidth, trim = 40 20 90 20, clip = true]{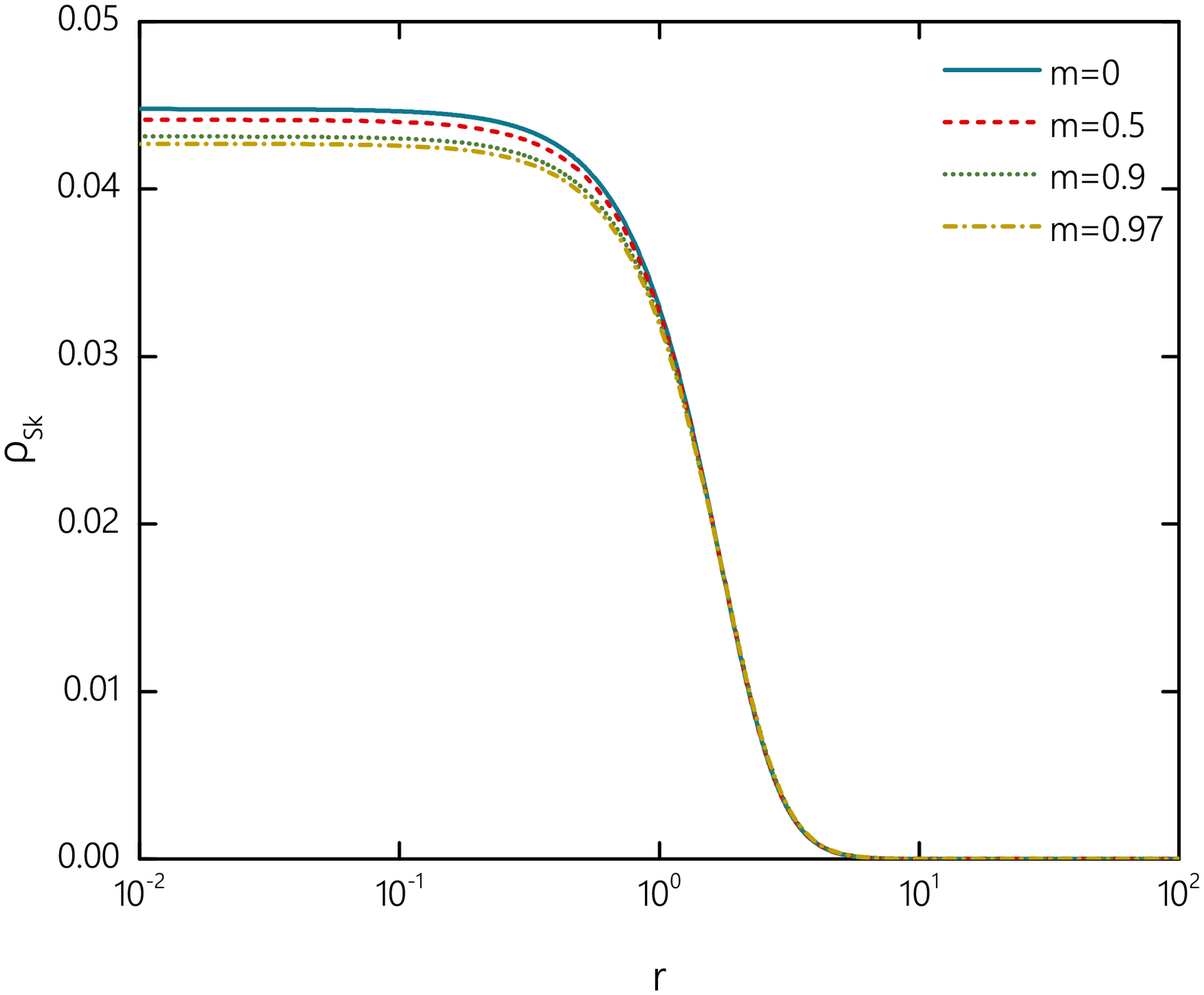}
\end{center}
\caption{\small
Components $v_1$ (solid line) and $u_2$ (dashed line) of the fermionic mode $A_0$ (upper left plot),
and components $v_2$ (solid line) and $-u_1$ (dashed line)
of the fermionic mode $A_0$ (upper right plot),
and the topological charge density distribution \re{charge} (bottom plot)
as functions of the radial coordinate
$r$ for some set of values of the fermionic mass $m$ at $g=1, \kappa_0=0.1$.}
\lbfig{fermmass}
\end{figure}

\begin{figure}[hbt]
\begin{center}
\includegraphics[width=.7\textwidth, trim = 60 20 80 50, clip = true]{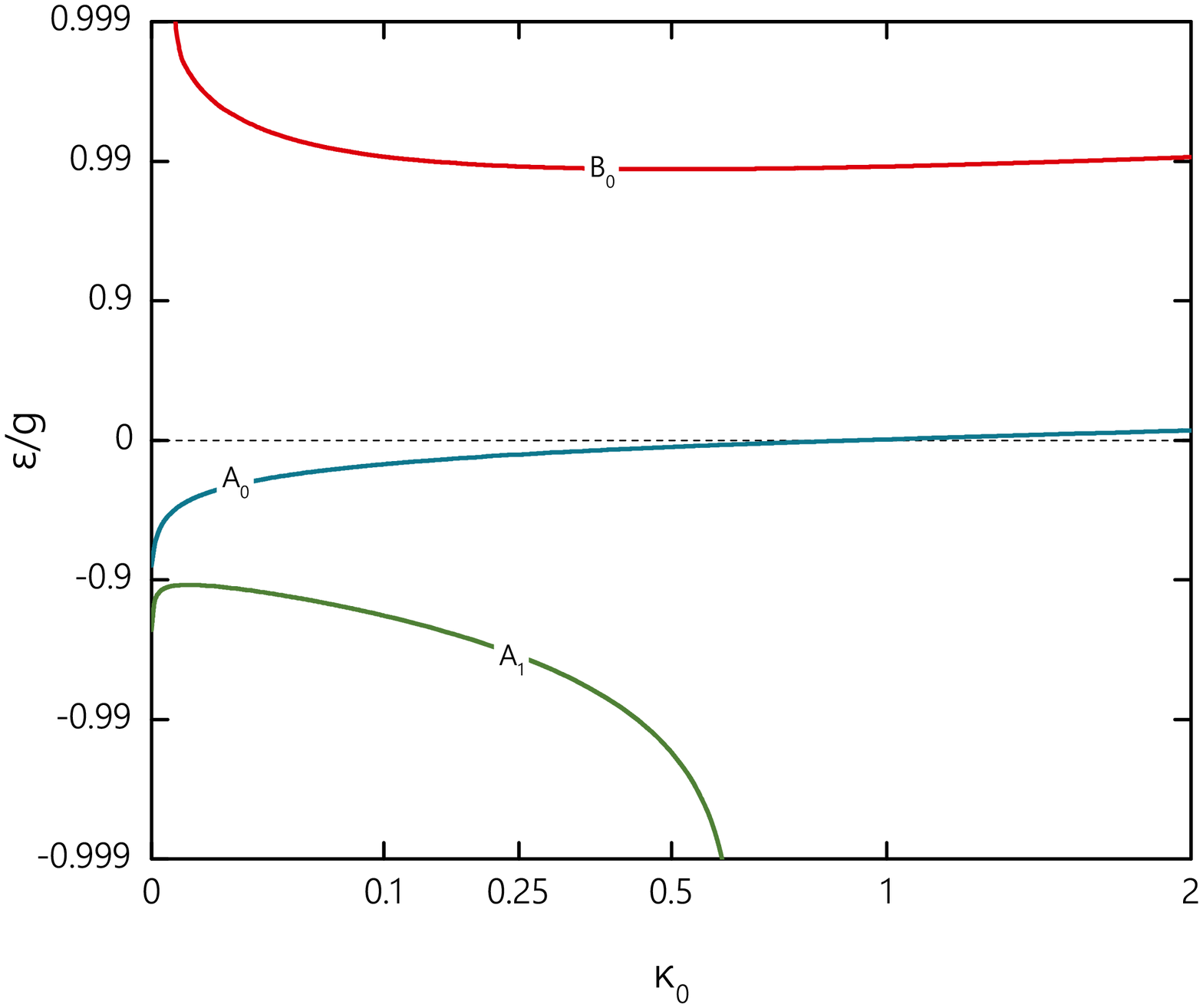}
\end{center}
\caption{\small
Normalized energy $\frac{\varepsilon}{g}$ of the localized fermionic states
as a function of the Skyrmion potential coupling $\kappa_0$
for several fermion modes at $m=0$ and $g=1$.}
\lbfig{engkappa}
\end{figure}

Next, we consider dependence of solutions on the value of the fermion mass $m$ for fixed $g$.
It is well known that for several model parameters the spectral flow behaviour can be realized.
In Fig.~\ref{modesm}, we present the plot which shows changing the mass parameter $m$ induces again a new spectral flow.
The energy of the nodeless mode $A_0$ is negative at $m=0$, and as increasing $m$
it crosses zero at around $m\approx 0.97$ (we also supply the zoomed subplot in Fig.~\ref{modesm}).
The spectral flow is more explicit as the coupling $g$ becomes stronger.
The energy of the localized fermionic states is restricted as $| \varepsilon | < |g-m| $.
The excitations of both types, $A_k$ and $B_k$ are delocalizing at some critical values of the fermion mass $m$.
On the other hand, decrease of the coupling constant $g$ also leads to delocalization of the fermionic modes, only massless
$m=0$ quasi zero mode $A_0$ remains as $g \ll 1$, see Fig.~\ref{modeslim}.


In Fig.~\ref{fermmass} we plot the topological charge density distribution
and the fermionic fields, for some set of values of the fermion mass  $0\leq m< 1$.
An interesting observation is that for large values of the parameter $m$ additional nodes may appear in the
fermionic field profile functions, thus the issue of classification of the modes, based on the number of nodes becomes
more subtle. For example, for $A_0$ mode with $\kappa_0=0.1$ additional node of $v_1$ function appears at $g=1, m\sim 0.95$, see
Fig.~\ref{fermmass}, left upper plot. Increase of the coupling $g$ makes this effect more explicit.




Let us consider how the fermionic modes are affected by the variation of
the Skyrmion mass parameter $\kappa_0$, as the coupling constant $g$ remains fixed.
In Fig.~\ref{engkappa} we display the normalized fermion energy in units of $g$  as a function of the coupling
$\kappa_0$ at $g=1$. As it is seen in Fig.~\ref{modeslim}, in this case there are two localized modes of type $A$ and one mode
of type $B$, the number of localized fermionic states increases for larger values of $g$.
First, we observe that increase of the "pion mass" parameter $\kappa_0$ also causes the spectral flow,
the energy of the nodeless mode $A_0$, which is negative at $\kappa_0=0$,
increases and crosses zero at some critical value of $\kappa_0$. Further increase of  $\kappa_0$ drives the eigenvalue $\epsilon$
towards positive energy continuum, it approaches it in the limiting case $\kappa_0 \to \infty$.
In a contrast, the energy of
the $A_1$ mode is decreasing, it approaches the negative continuum and decouples at $\kappa_0 \sim 0.61$.
The energy of the mode $B_0$,
which arise from the positive continuum at some small value of $\kappa_0$, is  initially decreasing, it has a minimum at some
value of the "pion mass" parameter. As $\kappa_0$ continue to grow, the corresponding eigenvalue starts to increase, it tends
to the positive energy continuum at $\kappa_0 \to \infty$.

\begin{figure}[hbt]
\begin{center}
\includegraphics[width=.45\textwidth, trim = 40 20 90 20, clip = true]{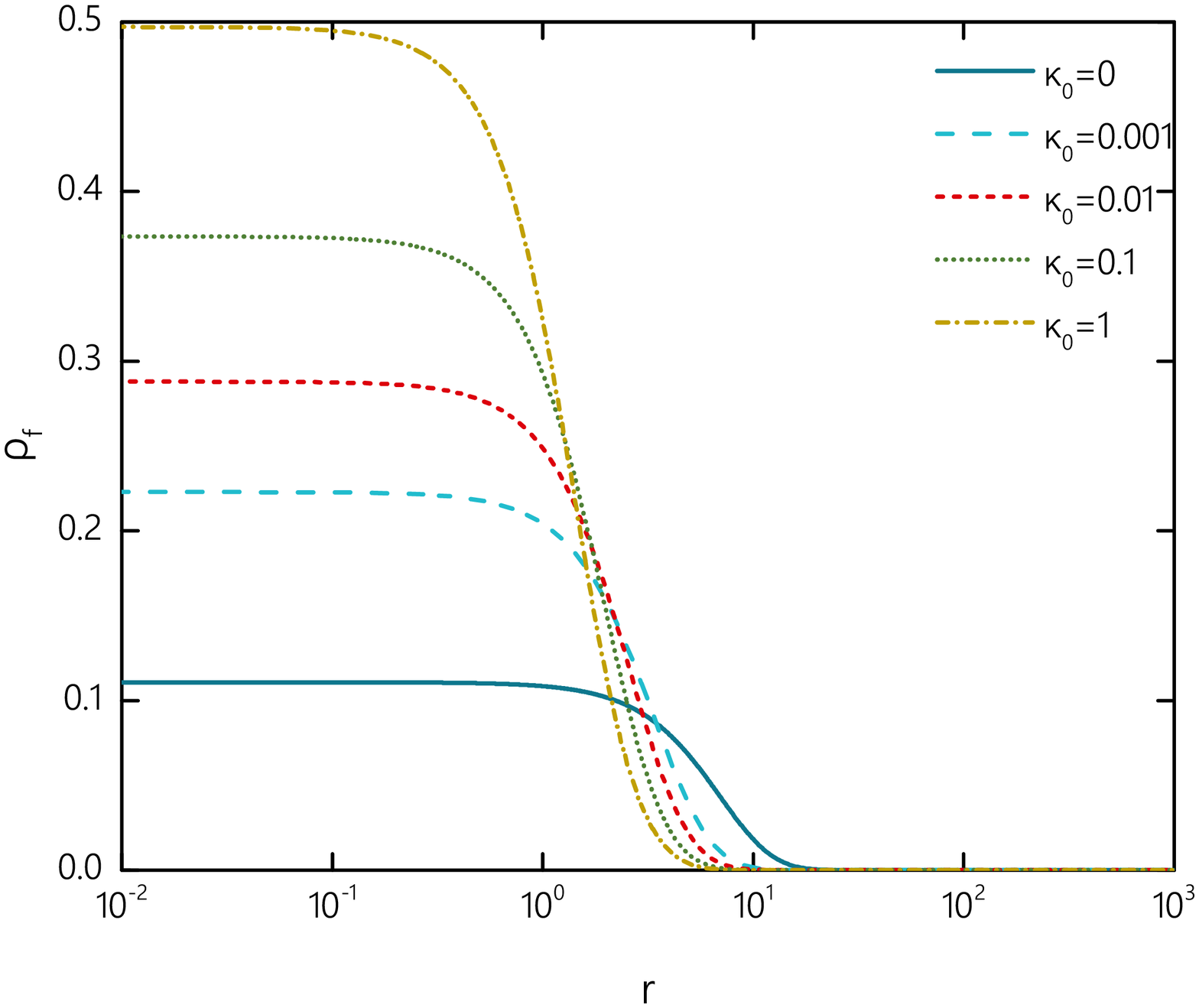}
\includegraphics[width=.45\textwidth, trim = 40 20 90 20, clip = true]{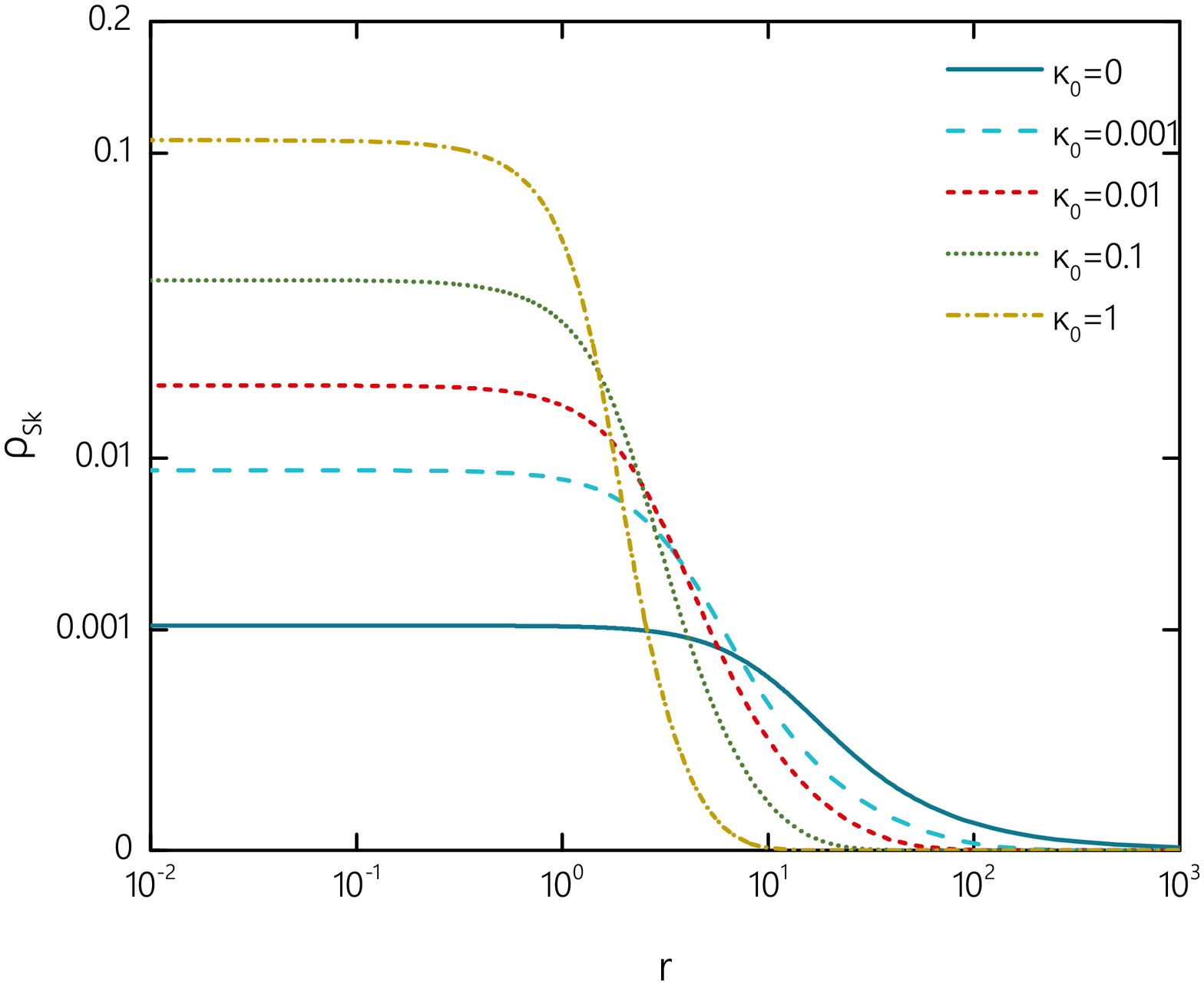}
\end{center}
\caption{\small
Distributions of the fermionic density of the localized mode $A_0$ (left plot) and the topological density of the Skyrmion
coupled to this mode (right plot) as functions of the radial coordinate
$r$ for some set of values of the coupling $\kappa_0$ at $m=0$ and $g=1$.}
\lbfig{profkappa}
\end{figure}

Note that the modes of type $A$ remain localized on the baby Skyrmion in the limiting
case $\kappa_0=0$. Indeed, our numerical results show that
coupling to the fermionic mode may stabilize the configuration, then
the strength of the coupling $g$ yields the characteristic scale of the soliton.
In Fig.~\ref{profkappa} we plotted the corresponding profiles of the fermionic density $\rho$ and the topological density
distribution for some set of values of the parameter $\kappa_0$.


Now, let us consider the radially excited modes, localized by the Skyrmion.
Numerical evaluation shows that, as a result of backreaction, the coupling to
higher fermionic modes yield much stronger
deformations of the Skyrmion than the coupling to the modes $A_0$ and $B_0$, which we discussed above.
However, since the corresponding numerical errors rapidly grows,
we restrict our consideration to the first a few $A_k$  modes.

As an example, in Fig.~\ref{radexkapp0g} we present the
radial distributions of the fermionic density of the localized $A_0$, $A_1$ modes
and the topological charge density of the baby Skyrmion, coupled to these modes.
\begin{figure}[hbt]
\begin{center}
\includegraphics[width=.45\textwidth, trim = 40 20 90 20, clip = true]{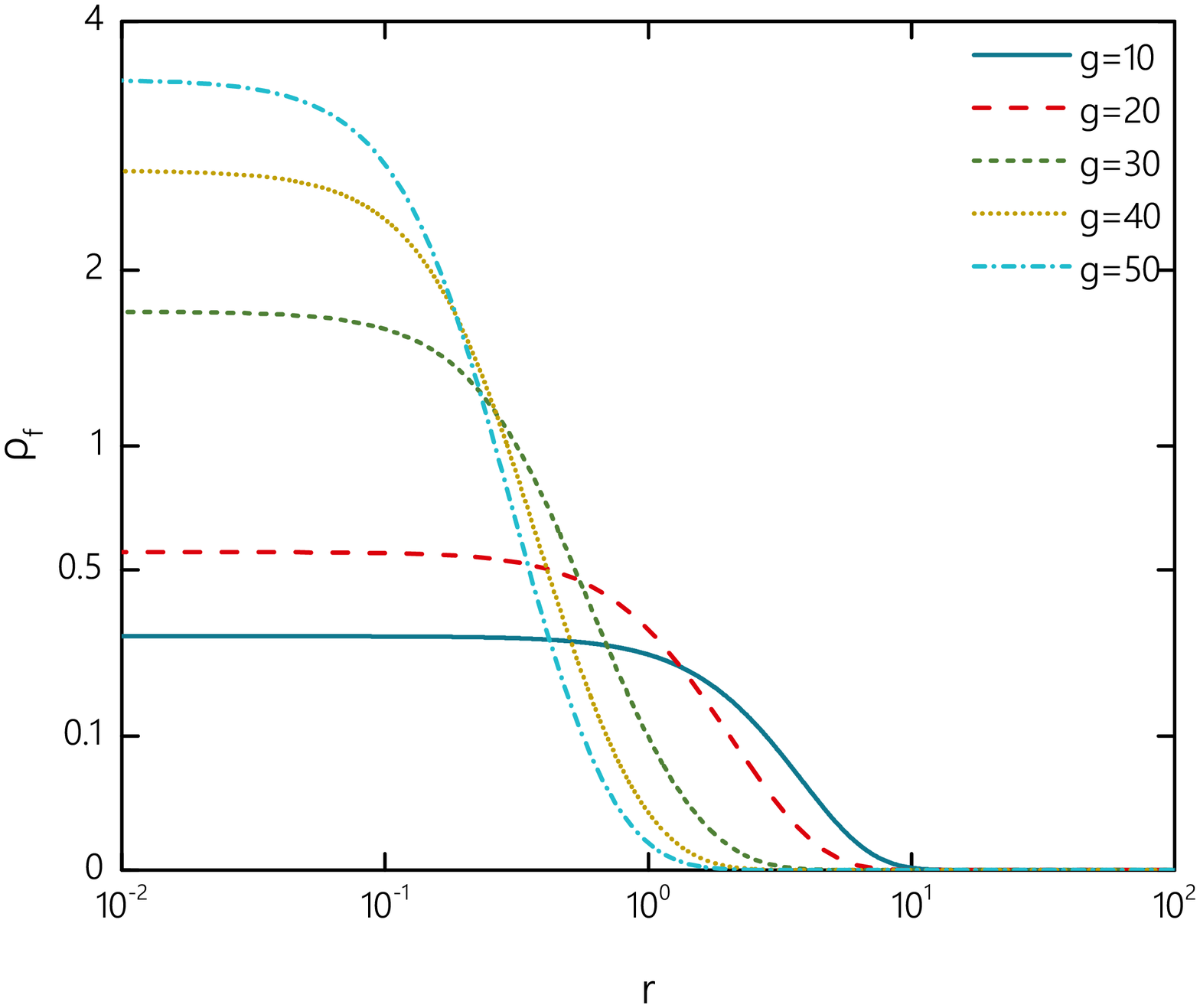}
\includegraphics[width=.45\textwidth, trim = 40 20 90 20, clip = true]{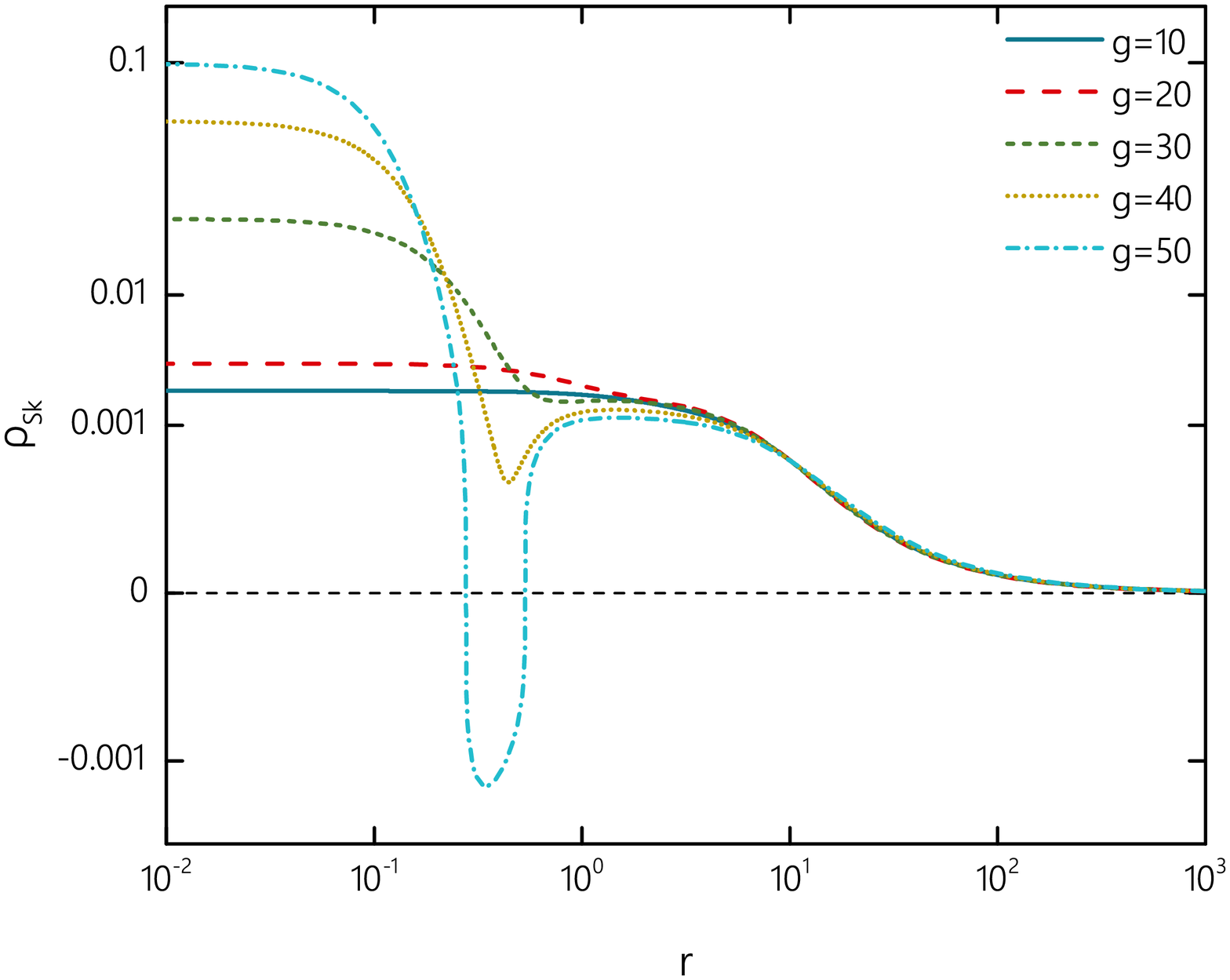}
\includegraphics[width=.45\textwidth, trim = 40 20 90 20, clip = true]{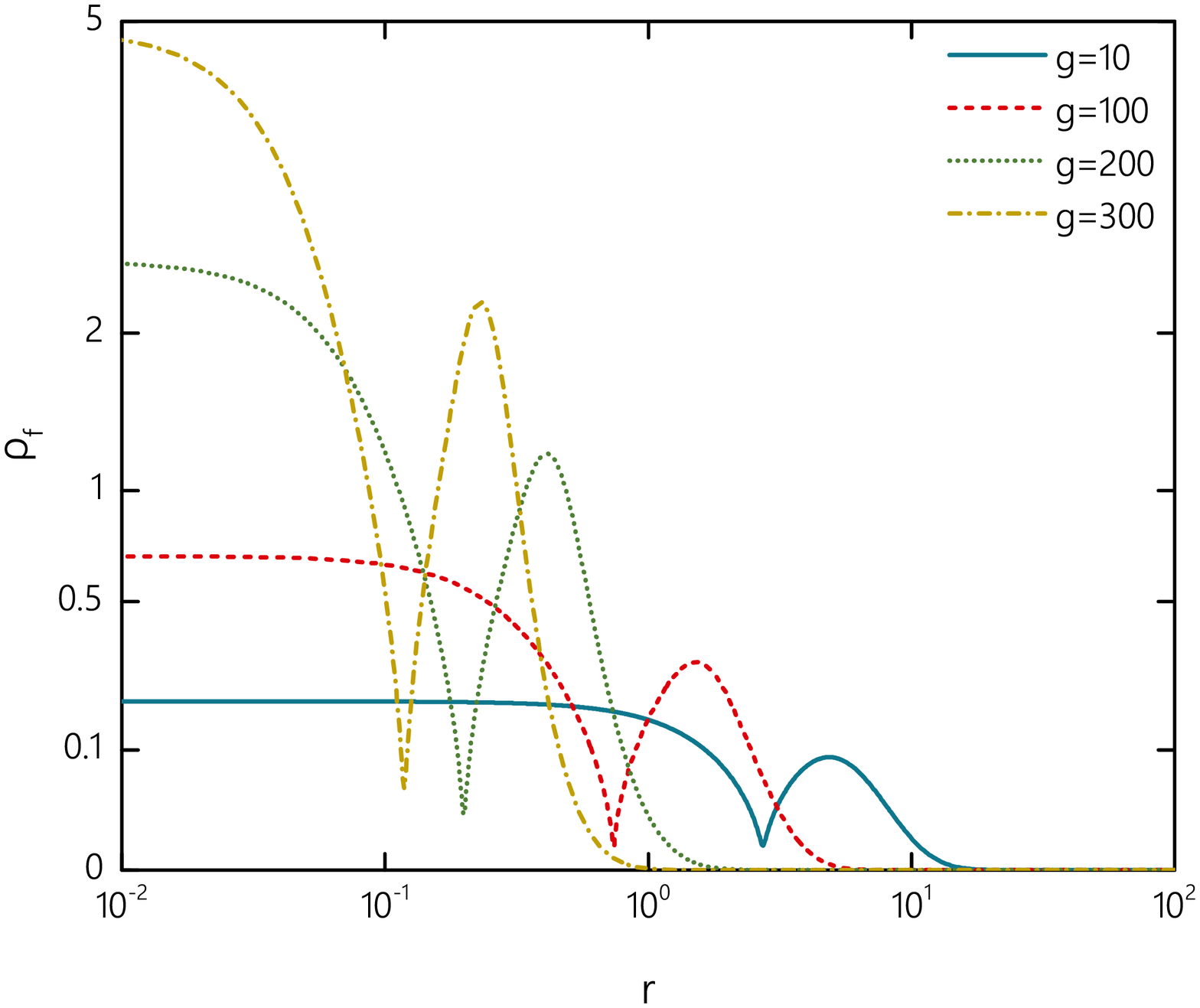}
\includegraphics[width=.45\textwidth, trim = 40 20 90 20, clip = true]{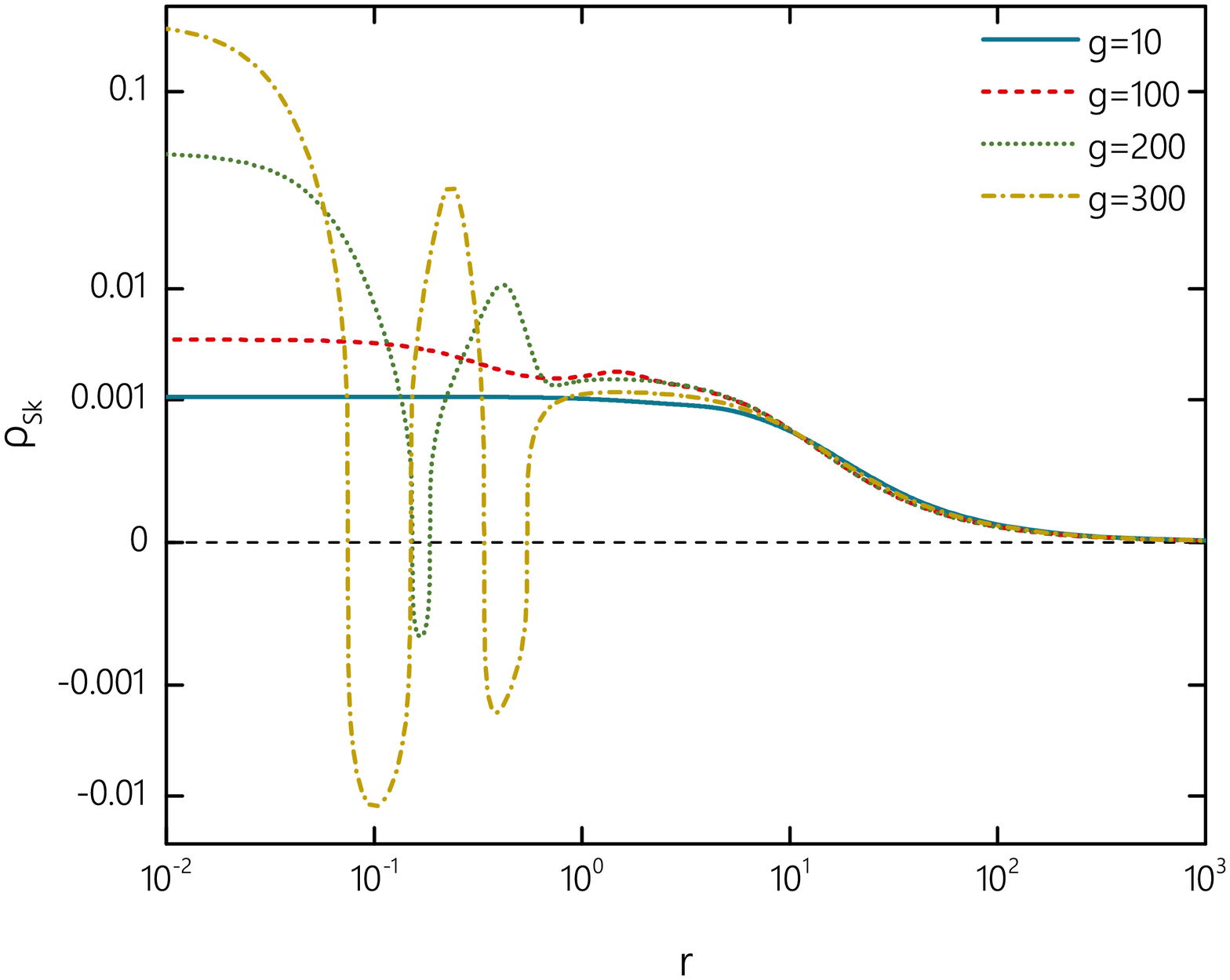}
\end{center}
\caption{\small
Distributions of the fermionic density (left plots) of the localized modes $A_0$ (upper plots) and $A_1$ (bottom plots)
and the topological density of the Skyrmion
coupled to these modes (right plots) as functions of the radial coordinate
$r$ for some set of  values of the coupling  $g$ at $m=0$ and $\kappa_0=0$.}
\lbfig{radexkapp0g}
\end{figure}

First, we observe that coupling to the higher modes strongly deforms the Skyrmion,
see Fig.~\ref{radexkapp0g}, right plots. In the strong coupling limit the profile function of the soliton is
no longer monotonically decreasing with $r$, a node of $f(r)$
appears and large amplitude oscillations of the profile function are observed. This effect becomes much more
explicit for small values $\kappa_0$. Physically, the oscillations of the topological charge density in
the coupled fermion-Skyrmion system may be
interpreted as production of the strongly bounded Skyrmion-anti-Skyrmion pair, thus the total topological charge of the
configuration does not change. However, the fermionic $A_1$ mode is now coupled to the concentric
multi-Skyrmion like configuration, it can be thought of as decomposed into the individual $A_0$
modes, localized by each of the constituents, see Fig.~\ref{radexkapp0g}, left plots.

\begin{figure}[t]
\begin{center}
\includegraphics[width=.45\textwidth, trim = 40 20 90 20, clip = true]{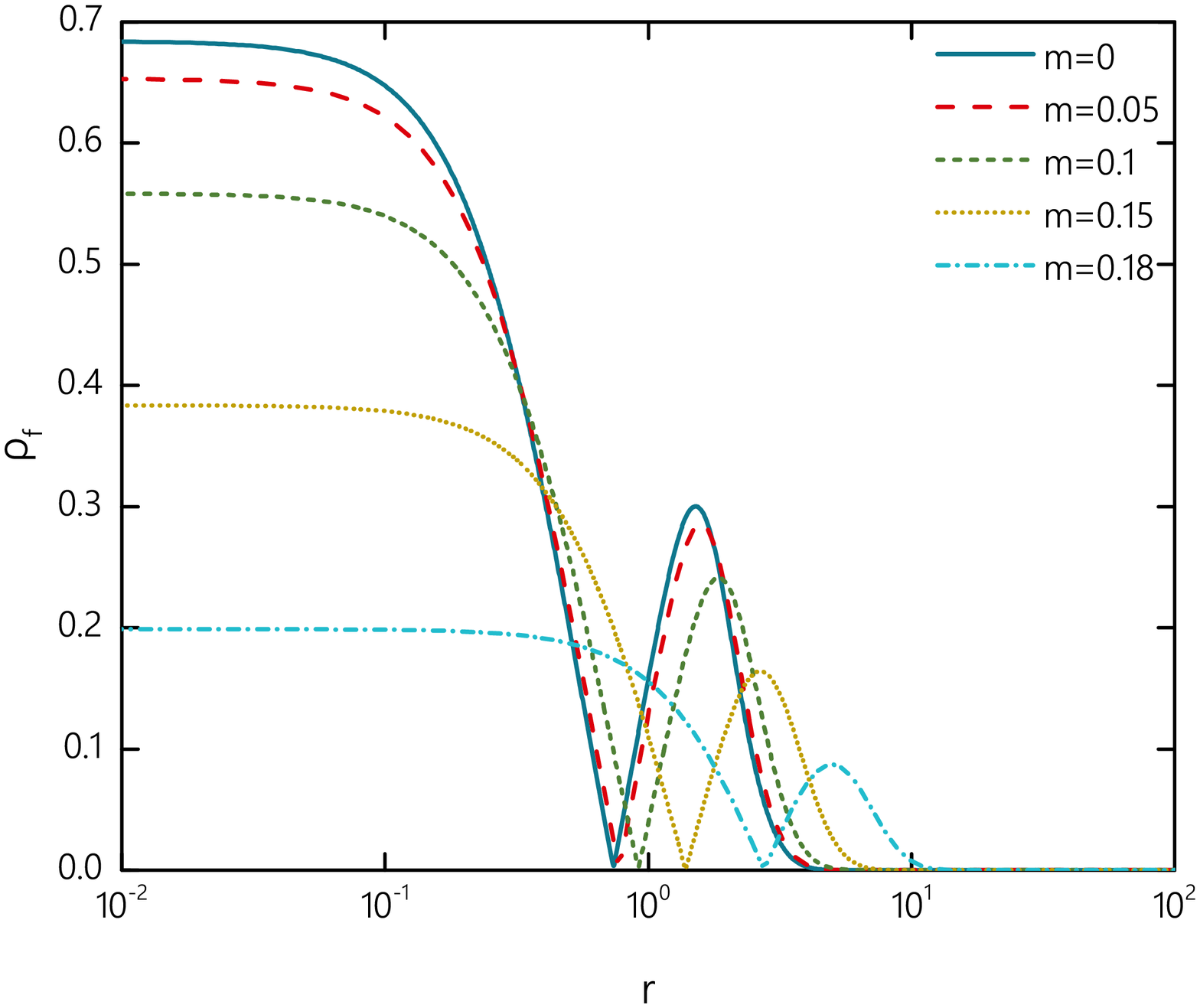}
\includegraphics[width=.45\textwidth, trim = 40 20 90 20, clip = true]{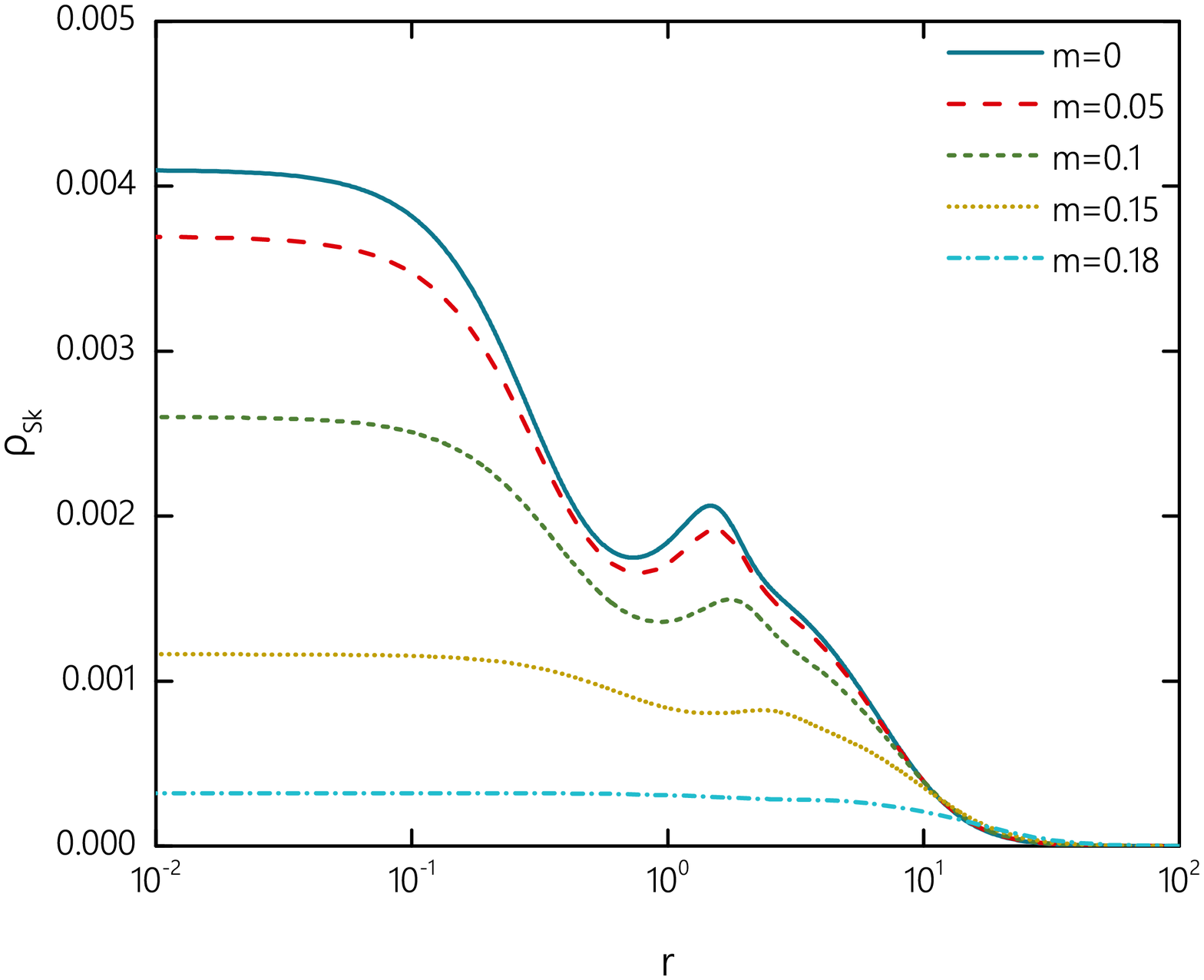}
\end{center}
\caption{\small
Distributions of the fermionic density of the localized mode $A_1$ (left plot) and the topological density of the Skyrmion
coupled to this mode (right plot) as functions of the radial coordinate
$r$ for some set of values of mass of fermions $m$ at $g=100$ and $\kappa_0=0$.}
\lbfig{fig10}
\end{figure}

Variation of the fermion mass parameter $m$ also affects the structure of the solutions. As a particular example, we considered
the dependency of the fermionic density of the localized mode $A_1$  and the topological density of the Skyrmion
coupled to this mode, for a fixed value of the coupling constant $g$, see Fig.~\ref{fig10}. Increase of the fermion mass
smooths out the spatial distribution of the densities, also the characteristic size of the configuration increases.

Finally, let us note that there can be several fermionic modes localized by the soliton. For low values of coupling $g$ topological and fermionic density profiles are almost unchanged relative to the case of localization of single fermion, though for higher $g$ deformations appears to be stronger. As a particular example, in  Fig.~\ref{fig11} we represent the $A_0$ and $A_1$ modes localized by the Skyrmion at $g=100$, this is the configuration with filling factor 2.

\begin{figure}[t]
\begin{center}
\includegraphics[width=.45\textwidth, trim = 40 20 90 20, clip = true]{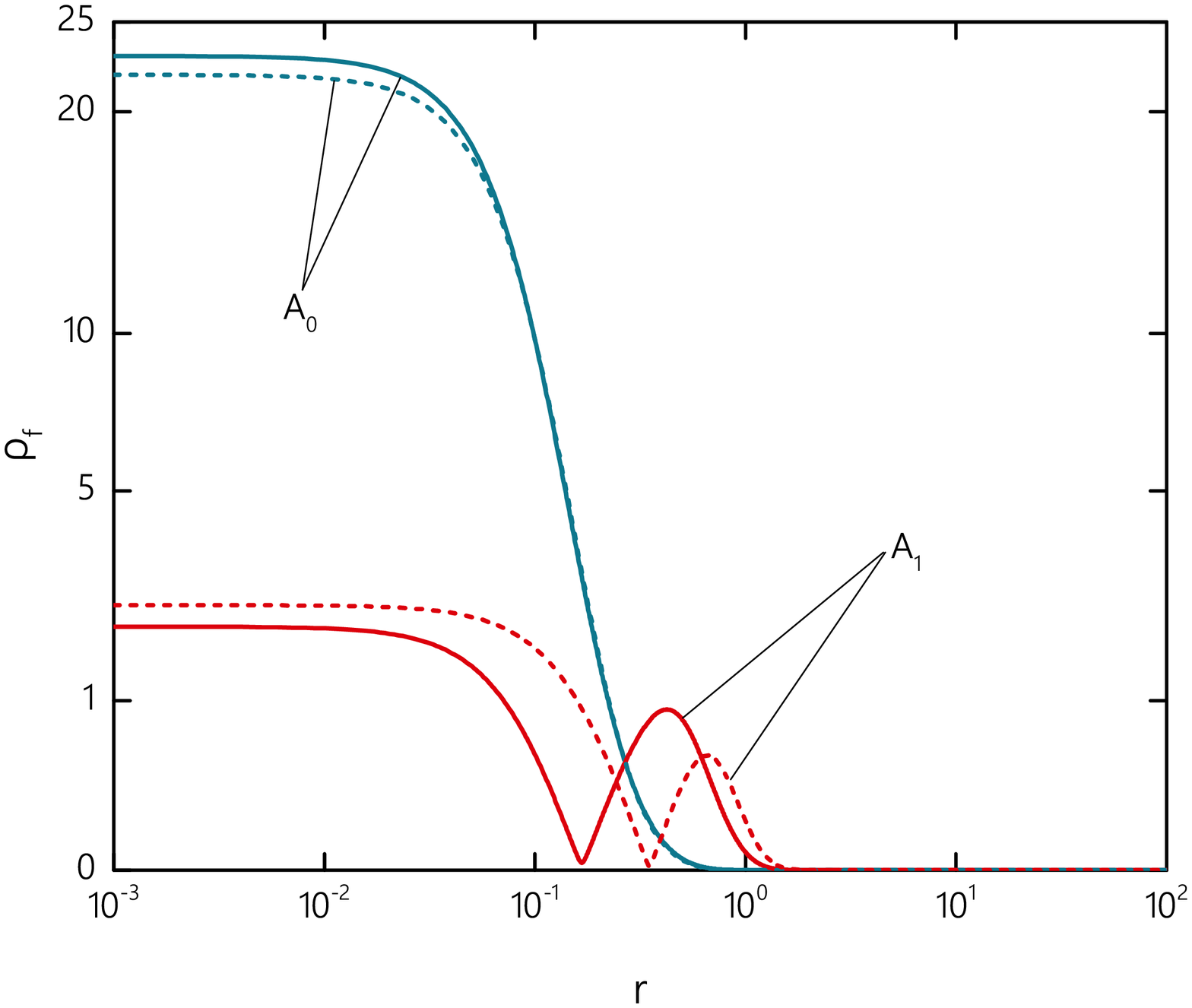}
\includegraphics[width=.45\textwidth, trim = 40 20 90 20, clip = true]{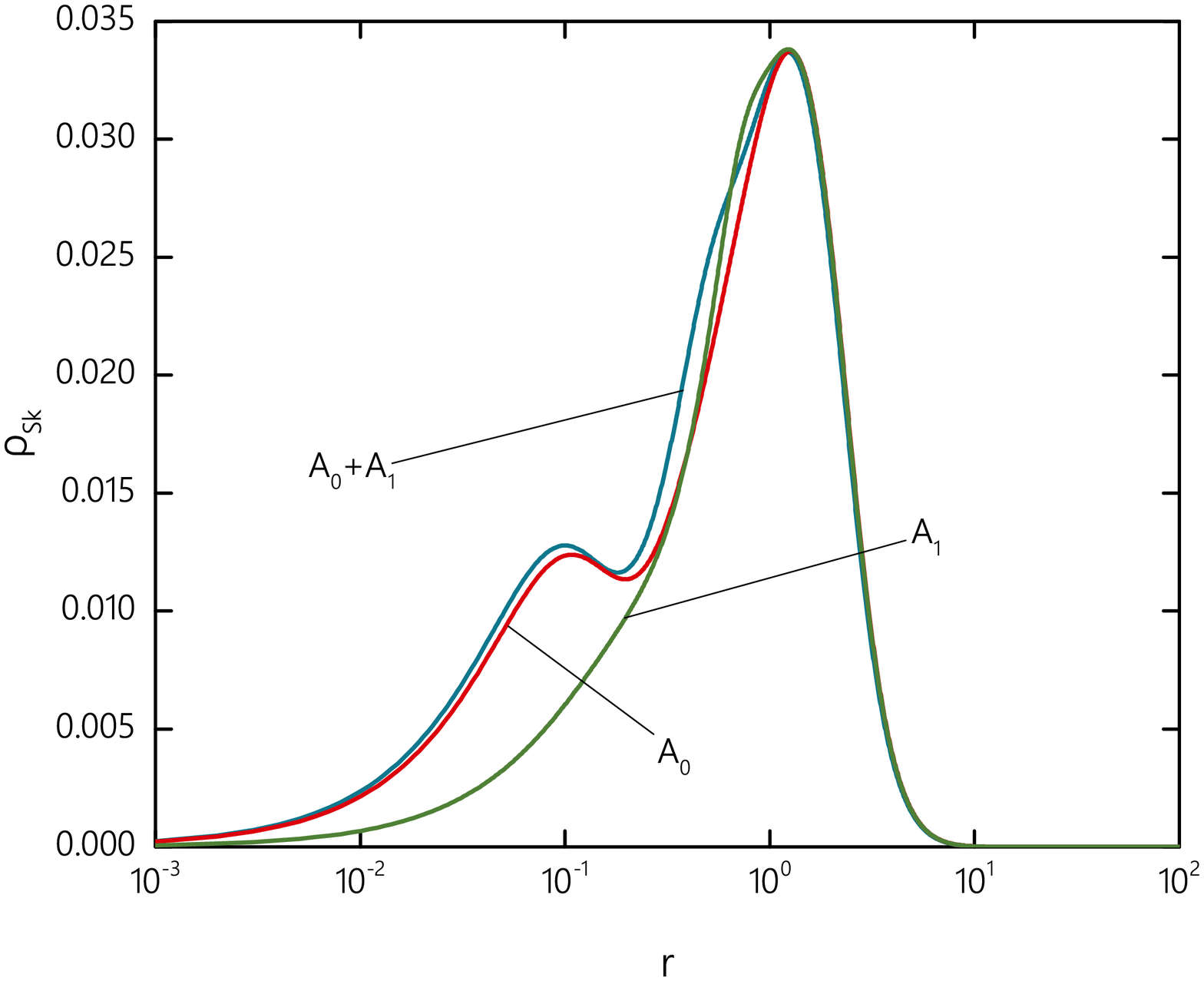}
\end{center}
\caption{\small
Distributions of the fermionic density of the localized modes $A_1$ and $A_2$ (left plot)
and the topological density of the Skyrmion
coupled to these modes (right plot) as functions of the radial coordinate
$r$  at $g=100$ and $\kappa_0=0$. The solid line corresponds to the state with filling factor 2, dashed lines correspond to
the states with filling factor 1.}
\lbfig{fig11}
\end{figure}

\section{Summary and conclusions}
The objective of this work is to investigate the effects of backreaction of the fermions coupled to the planar Skyrmions.
We found that there are two different types of the fermionic modes, localized on the Skyrmion, in particular
there is only one zero crossing bounded nodeless mode. Other modes, which are linked to the positive and negative continuum,
do not cross zero in agreement with the index theorem. Decrease of the coupling constant leads to delocalization of the
fermionic modes, only massless quasi zero mode remains in the weak coupling limit. We show that the coupling to the fermions
may stabilize the soliton in the limiting case of vanishing potential, then the coupling strength defines the characteristic
size of the Skyrmion. Considering the strong coupling limit we found that the coupling to the fermionic modes strongly deforms
the Skyrmion, in particular we observe production of tightly bounded Skyrmion-anti-Skyrmion pair.

The work here should be taken further by considering the multisoliton solution in the planar Skyrme model with fermionic fields.
The additional long-range interaction mediated by the fermions, localized on the solitons, may strongly affect the usual pattern
of interaction between the Skyrmions. Another direction can be related with investigation of properties of fermions localized on
solitons in the baby Skyrme model with the Dzyaloshinskii-Moriya interaction term. We hope to address these problems in our future
work.

\section*{Acknowledgements}
N.S. and Ya.S. gratefully acknowledge support from the
Foundation for the advancement of theoretical physics and Mathematics "BASIS".
Ya.S. would like to thank Steffen Krusch for many discussions which initiated this study long time ago,
and for his valuable most recent remarks.

\end{document}